\documentclass[twocolumn,secnumarabic,amssymb, nobibnotes, aps, prl]{revtex4-2}
\usepackage{amssymb}
\usepackage{amsthm}
\usepackage{graphicx}
\usepackage{amsmath}
\usepackage{color}

\begin{document}


\title{Realization of high-fidelity perfect entanglers between remote superconducting quantum processors}

\author{Juan Song$^{1,2,3}$}\thanks{these authors contributed equally to this work}
\author{Shuang Yang$^1$}\thanks{these authors contributed equally to this work}
\author{Pei Liu$^4$}\thanks{these authors contributed equally to this work}
\author{Hui-Li Zhang$^1$}
\author{Guang-Ming Xue$^1$}
\author{Zhen-Yu Mi$^1$}\email{mizy@baqis.ac.cn}
\author{Wen-Gang Zhang$^1$}\email{zhangwg@baqis.ac.cn}
\author{Fei Yan$^1$}
\author{Yi-Rong Jin$^1$}\email{jinyr@baqis.ac.cn}
\author{Hai-Feng Yu$^{1,5}$}

\affiliation{$^1$Beijing Academy of Quantum Information Sciences, Beijing 100193, China}
\affiliation{$^2$Institute of Physics, Chinese Academy of Science, Beijing 100190, China}
\affiliation{$^3$University of Chinese Academy of Sciences, Beijing 101408, China}
\affiliation{$^4$State Key Laboratory of Low Dimensional Quantum Physics and Department of Physics, Tsinghua University, Beijing 100084, China}
\affiliation{$^5$Hefei National Laboratory, Hefei 230088, China}

\begin{abstract}

Superconducting qubits, a promising candidate for universal quantum computing, currently face limitations in chip size due to reproducibility, wiring complexity, and packaging modes. Distributed quantum modules offer a viable strategy for constructing larger quantum information processing systems, though universal quantum gate operations between remote qubits have yet to be realized. Here, we demonstrate high-fidelity perfect entanglers between two remote superconducting quantum devices over 30 cm distance, leveraging the standing-wave modes in the coaxial cable connecting them. We achieve cross-entropy benchmarking (XEB) fidelities of $(99.15 \pm 0.02)\%$ and $(98.04 \pm 0.04)\%$ for CNOT and CZ gates, respectively, which are more efficient and universal than existing state transfer or feedback-based protocols. This advancement significantly enhances the feasibility of universal distributed quantum information processing, essential for the future development of large-scale quantum systems.
\end{abstract}

\maketitle

\section{Introduction}
Superconducting quantum information processors have demonstrated their advantage in quantum computation and quantum simulation against classical computers, especially when the number of qubits exceed a certain scale. \cite{supermacy_Martinis, wu2021strong, surface_google, kim2023evidence, rosen2024implementingsyntheticmagneticvector}. Furthermore, to execute useful algorithms through quantum error correction, significantly larger quantum processors with $\sim50$ million qubits are necessary \cite{surface_theo}. It is almost impossible to fabricate such a large-scale quantum processor on a single chip due to the limitation of chip size, which currently allows for approximately $1000$ qubits. Meanwhile, the topological connectivity due to the planar geometry also leads to an increase in the number of qubits or the depth of the quantum circuit while implementing the quantum algorithm and quantum error correction codes which require high connectivity \cite{QEC}. Thus, distributed quantum computation attracts wide attention, which offers a viable solution for scaling up quantum processors by connecting multiple small-scale processors to facilitate large-scale quantum computation \cite{distributed_Lukin, largescale_luming, silicondie_rigetti, ARQUIN_architectures}. The key to implementing universal quantum computation in distributed quantum systems is to realize the essential element of the universal gate set — the two-qubit gate — between different processors \cite{barenco1995elementary}. Recent studies on the quantum state transfer (QST) process have demonstrated that employing superconducting transmission lines to link small processors is a potential way for scaling up superconducting quantum computation \cite{bidiractional_cleland, violateBell_zhong, quantumlink_walraff, Devoret_deterministic, deterministicentanglement_zhong, purification_zhong, loopholefree_walraff, remote_Chang, lowloss_zhong}. However, the flying-photon-mediated QST process cannot be directly integrated into quantum circuits since it is not a unitary two-qubit gate. Even the utilization of bi-directional flying photons to implement a SWAP gate does not result in a perfect entangler, thus precluding its use in forming a universal computation gate set \cite{perfectentangler_zhang}. Recent theoretical investigations have explored the cross-resonance (CR) effect between two remotely connected quantum devices \cite{remoteCR}. The CR effect is widely used to establish an entanglement due to its simplicity in implementation and insensitivity to flux noise, which is of great potential in recent quantum information processing research \cite{QV64_IBM, degroot_selectivedarkening, fullMW_regetti, chow_simpleAllMW, degroot_selectivedarkening_2, Corcoles_ProcessVerification, Patterson_CaliCRgate, Tripathi_OperationIntrinsicerrorbudget, ware_cross,Ku_supperssionZZ,high_fidelity,Malekakhlagh_firstprinciple_pra,megasan_effective_pra,mitchell_HardwareEfficient,Wei_CrosstalkCancellation,xu_zzfreedom}. It is noticed that the CNOT gate based on CR effect, as well as the similar stark-induced CZ gate \cite{mitchell2021hardware, wei2022hamiltonian}, can be employed to build a remote and universal quantum computation gate set.

In this work, we report the realization of direct two-qubit entangling gates between two separately packaged quantum processors. Unlike the feedback-based protocol \cite{deterministictele_zhong}, our remote gate exhibits both short duration ($204$ ns) and high fidelity ($99.15 \%$). We use XEB to benchmark the two-qubit gate and implement Bell-inequality test to confirm these unitary manipulations. Besides these features, our method can also enable a more complex connecting topology among qubits in a 2D structure compared to the previous inter-module approaches \cite{silicondie_rigetti}, leading to a more feasible error correction code realization \cite{QEC}. Such advancements provide a convenient and universal protocol to implement distributed quantum information processing, which paves the way for sophisticated large-scale quantum engineering in the future.

\section{Results}

\begin{figure}[htbp]
    \centering
    \includegraphics[]{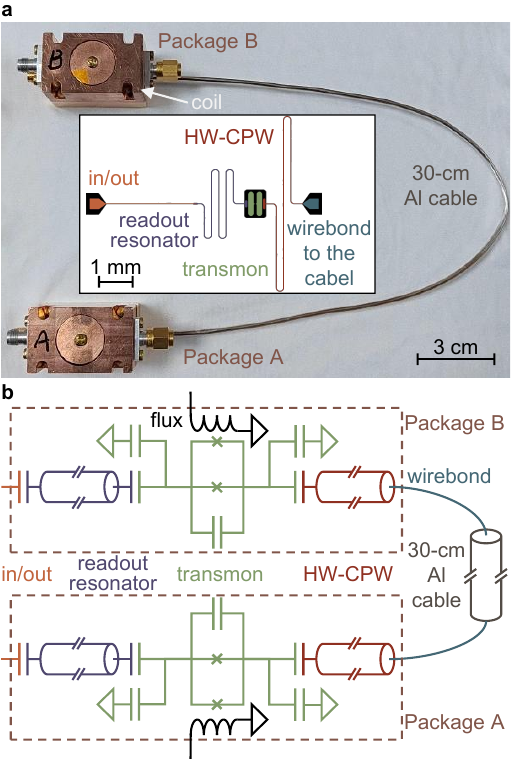}
    \caption{\textbf{A circuit schematic for the remotely coupled system.} \textbf{a}, Photograph of two packages connected by a superconducting coaxial cable. There is a coil on each copper sample box. A small coil is installed inside each package to provide external flux for tunning the frequency of the qubits. Inset in \textbf{a}, Layout of one chip which consists of a transmission line (orange), a resonator (purple), a transmon (green), and a half-wavelength co-planar waveguide (red). \textbf{b}, Circuit schematic.}
    \label{fig:cr_fab}
\end{figure}

\begin{figure}[htbp]
    \centering
    \includegraphics{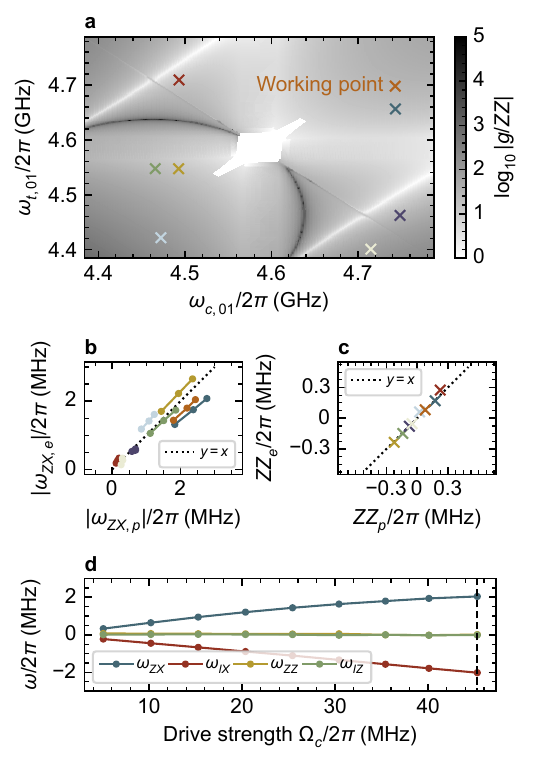}
    \caption{\textbf{Simulated results of cross-resonance effect in our device configuration.} \textbf{a}, The ratio of the effective coupling strength $g$ to the $ZZ$ strength when coupled via single mode $n=15$. The central white area is ignored in simulations since the frequencies of qubits and bus mode $n=15$ are close. \textbf{b}, Considered modes $n=14, 15, 16$ ($16$ is inferred), the effective CR strength $|\omega_{ZX}|$ is predicted by simulations, compared with the experimental $|\omega_{ZX}|$ at different frequencies for the two qubits and three drive strengths. \textbf{c}, The residual $ZZ$ strengths. The different working frequencies are labeled by corresponding colors in \textbf{a}, \textbf{b},  and \textbf{c}. \textbf{d}, The CR parameters as a function of driving strength at the working point. The dashed line is the selected drive strength.}
    \label{fig:CR_str}
\end{figure}

\begin{figure*}[htbp]
    \centering
    \includegraphics[]{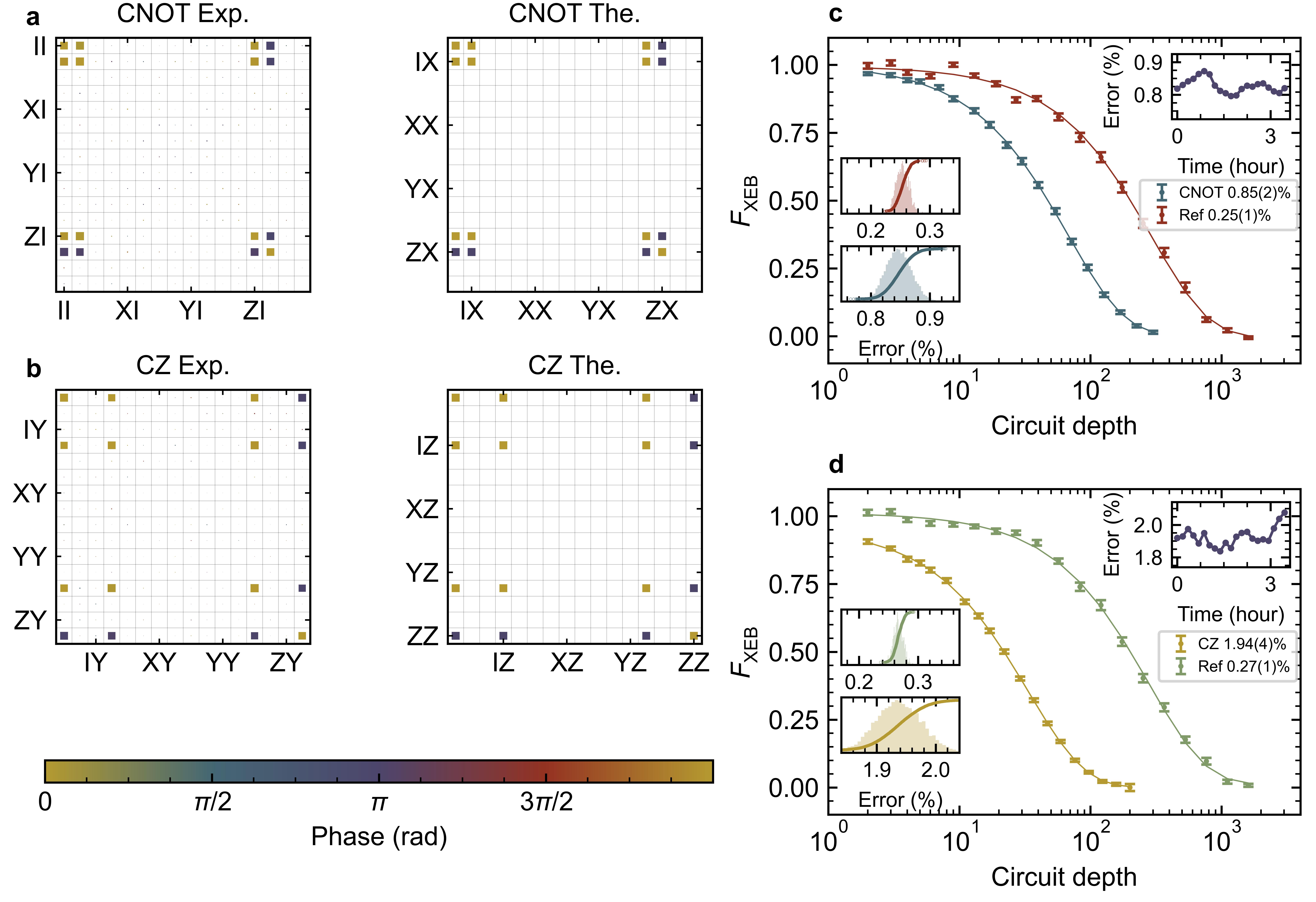}
    \caption{\textbf{Quantum process tomography and cross-entropy benchmarking}. \textbf{a}, The quantum process matrix of the experimental CNOT gate as well as its theoretical matrix with QPT fidelity $F=98.72\%$. \textbf{b}, The quantum process matrix of the experimental CZ gate as well as its theoretical matrix with QPT fidelity $F=96.01\%$. The elements of each complex matrix are shown by a colored square with its length indicating the ratio of the absolute value to $0.5$ and its color indicating the phase. Both gates are benchmarked by XEB shown in \textbf{c} and \textbf{d}, respectively. The left insets are the histograms of the reference and gate error rates calculated by bootstrapping. The right one is the time stability of the gate error rate.}
    \label{fig:XEB}
\end{figure*}

\begin{figure*}[htbp]
    \centering
    \includegraphics[]{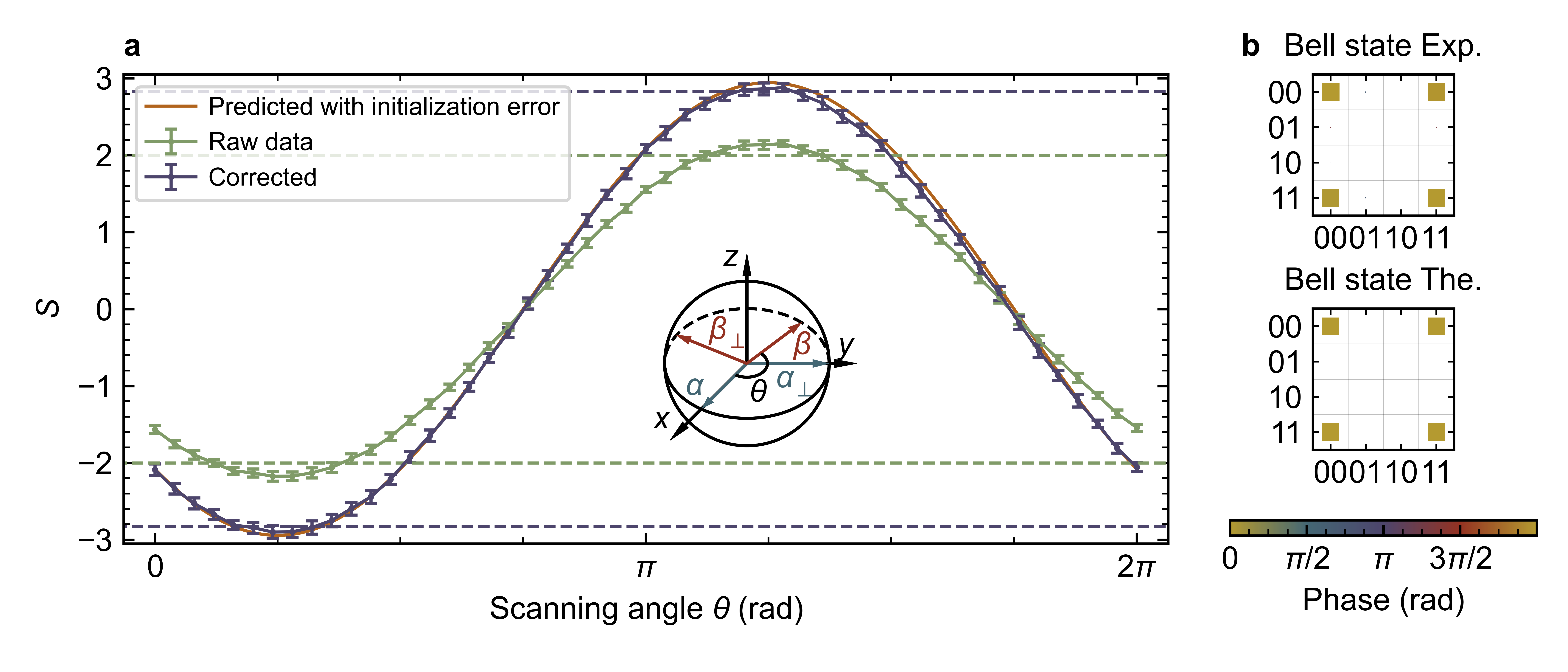}
    \caption{\textbf{Bell inequality measurement and Bell state preparation}. \textbf{a}, The CHSH correlation $S$ as a function of scanning angle $\theta$. The green dots indicate the data without readout correction while the purple ones indicate the data after readout correction. Classical and quantum limits are marked with horizontal dashed lines. The red line indicates the experimentally measured data predicted with initialization error. \textbf{b}, The experimental and theoretical density matrices of the Bell state $|\Phi^+\rangle$ with fidelity $F=99.14\%$. The length of each square is the ratio of the absolute value of each complex element to $1$.}
    \label{fig:bell}
\end{figure*}

Our device shown in FIG. \ref{fig:cr_fab} consists of two packages. Each package contains one transmon qubit, the frequency of which can be tuned by the magnetic flux generated by the coil mounted inside the package, beneath the quantum chip. The transmon qubit is capacitively coupled to a readout resonator, as well as a half-wavelength co-planar waveguide (HW-CPW). The other end of the HW-CPW is bonded to a 30-cm-long aluminum (Al) coaxial cable with superconducting Al wire. The bonding point sites on the node of the current standing wave mode. The two packages are mounted on the mixing chamber cold plate of a dilution refrigerator, shielded with a cryo $\mu$-metal can to  reduce the impact of magnetic flux fluctuation (See Methods).

The system Hamiltonian is
\begin{equation}
\begin{aligned}
H&=\sum_{i=c, t}\omega_ia_i^\dagger a_i+\frac{\alpha_i}2a_i^\dagger a_i^\dagger a_ia_i+\sum_n\omega_na_n^\dagger a_n\\
&+\sum_ng_{c,n}\left(a_c^\dagger+a_c\right)\left(a_n^\dagger+a_n\right)\\
&+\sum_n(-1)^ng_{t,n}\left(a_t^\dagger+a_t\right)\left(a_n^\dagger+a_n\right)
\end{aligned}\label{eq:sys_H}
\end{equation}
with transmons described by Duffing oscillators. We measure the spectra of both qubits as a function of the magnetic flux bias. Several equidistant anti-crossing points indicate the interaction between the qubits and the standing wave mode of the transmission line (i.e. the bus mode). By fitting the spectra close to the anti-cross point, we extract $\omega_n$ the frequency of the $n$th standing wave mode, and $g_{c(t), n}$ the coupling strength to the qubits (See Methods). Given that all the related bus modes are on their ground state, the effective coupling strength $g$ and the residual $ZZ$ interaction between two qubits are what we concerned. A simple simulation result with only one bus mode ($n=15$) considered is shown in FIG. \ref{fig:CR_str}a (parameter details and other simulation results are in Methods). Assisted by the numerical simulation results, we choose working frequencies of the control and target qubits with weak $ZZ$ interaction strength and strong coupling strength, as well as weak impact of the spectator bus modes. Here we emphasize that the flux bias is used solely for determining the working point. Therefore, we treat our qubits as frequency-fixed and do not apply any flux pulse throughout the experiment.

The CR gate is one of the most widely employed gates to generate entanglement in frequency-fixed qubit system \cite{QV64_IBM, remoteCR}. To simplify, we write the effective two-qubit CR drive Hamiltonian in terms of $IX$, $IY$, $IZ$, $ZX$, $ZY$, and $ZZ$,
\begin{equation}
\begin{aligned}
H_{\rm eff} &= \omega_{IX} IX+\omega_{IY} IY+\omega_{IZ} IZ\\
   &+\omega_{ZX} ZX+\omega_{ZY} ZY+\omega_{ZZ} ZZ,
\end{aligned}\label{eq:CR_eff}
\end{equation}
where $X$, $Y$ and $Z$ are the Pauli matrices and $I$ is the unitary matrix. Using the Hamiltonian tomography method in Ref \cite{CRparameters_sheldon}, we can extract the six parameters in Eq. (\ref{eq:CR_eff}). Then, we vary the driving amplitude $\Omega_c$ and obtain these Hamiltonian parameters as a function of the driving amplitude, as shown in FIG. \ref{fig:CR_str}d. Together with several different biases of the control and target qubits, we extract CR parameter $|\omega_{ZX}|$ and residual $ZZ$ strength from the experimental data. The result and its comparason with the simulated interaction strengths are shown in FIG. \ref{fig:CR_str}b and c. It is found that the predicted results considered three bus modes agree well with the experimental results, indicating that the coupling between two qubits is contributed by several related bus modes simultaneously (See Methods).

We notice that ${\rm CX}=\exp{\left[-i\pi(ZX-IX)/4\right]}$ and ${\rm CNOT}=e^{-i\pi/4}\cdot{\rm CX}\cdot\exp{(-i\pi ZI/4)}$. Then, we finely tune the drive parameters to cancel the unwanted $IY$, $ZY$, $IZ$ terms, and set $\omega_{IX}=-\omega_{ZX}$, leading to a CNOT gate with ignorable $ZZ$ interaction. Meanwhile, when the drive frequency is properly detuned, the $X$, $Y$-like terms vanish under the rotating wave approximation, and the $ZZ$ term becomes dominant. This stark-induced $ZZ$ interaction can be employed to realize a CZ gate. The details of gate calibration method can be found in Methods.

Quantum process tomography (QPT) and XEB are utilized to evaluate the performance of the calibrated CNOT and CZ gates. The QPT process matrix $\chi$ is obtained and compared to the ideal matrix $\chi_{\rm CNOT}$ and $\chi_{\rm CZ}$, as shown in FIG. \ref{fig:XEB}a and b. The process fidelity is measured to be $F_{\rm CNOT}={\rm Tr}(\chi\cdot\chi_{\rm CNOT} ) = 98.72\%$ and $F_{\rm CZ}= 96.01\%$. The state preparation and measurement (SPAM) error is not ruled out in the QPT experiment. Furthermore, the SPAM-independent XEB method is implemented \cite{blueprint_Martinis}. By fitting the decay rate of the XEB fidelity as the number of CNOT (CZ) gates, shown in FIG. \ref{fig:XEB}c and d, the average fidelity of our CNOT (CZ) gate is estimated to be $(99.15 \pm 0.02)\%$ ($(98.04 \pm 0.04)\%$) with gate duration of $204$ ns ($430$ ns). Considering the relaxation and dephasing time of our qubits, we believe the gate infidelity mainly comes from qubit decoherence.

Finally, we show the ability to establish a remote entanglement between the two packages by testing the violation of the Bell inequality. We prepare the system onto one of the Bell states $|\Phi ^+\rangle =(|00\rangle +|11\rangle) / \sqrt{2} $, which is shown in FIG. \ref{fig:bell}b with fidelity of $99.14\%$. We then perform the Clauser–Horne–Shimony–Holt (CHSH) form of Bell inequality test \cite{CHSH} on our remotely entangled state, with the result shown in FIG. \ref{fig:bell}a. The control qubit is measured along $\alpha =x$ or $\alpha_{\perp}=y$. Simultaneously, the target qubit is measured along $\beta$ or $\beta_{\perp}$. The angle between $\alpha$ and $\beta$ is scanned from $0$ to $2\pi$. Without readout correction, at $\theta = 5\pi/4$, the maximum CHSH correlation $S=2.15 \pm 0.04$ exceeds the maximum classical value 2 by 3.8 standard deviations. After readout correction, the maximum correlation $S=2.88\pm 0.05$. The correlation is slightly higher than the quantum limit $2\sqrt{2}$. This error mainly comes from the initialization error on two qubits (See Methods).

\section{Conclusion}

We investigate the coupling form of a remote-coupled system and precisely tune the drive parameters to realize CNOT and CZ gates between two separately packaged quantum processors. The average fidelity is $(99.15 \pm 0.02)\%$ for CNOT and $(98.04 \pm 0.04)\%$ for CZ, characterized through the XEB method. In addition to its high fidelity, our hardware-efficient approach does not require any post-selection or feedback, making it highly data-efficient, low-latency, and engineering-friendly. These advancements represent a significant step toward scalable quantum information processing, providing a robust foundation for developing sophisticated quantum algorithms and error correction techniques necessary for practical, large-scale quantum systems. Moreover, we also realize the non-consistency between the $XX$-coupling model and relativity, which guides a direction for deep investigation in future studies. We believe the related theoretical and experimental progress may offer new insight to quantum information.

\textit{Note added} --- We noticed another work on realizing remote two-qubit gate on superconducting quantum processors recently \cite{mollenhauer2024high}.

\bibliography{reference_nature}

\begin{thebibliography}{51}%
\makeatletter
\providecommand \@ifxundefined [1]{%
 \@ifx{#1\undefined}
}%
\providecommand \@ifnum [1]{%
 \ifnum #1\expandafter \@firstoftwo
 \else \expandafter \@secondoftwo
 \fi
}%
\providecommand \@ifx [1]{%
 \ifx #1\expandafter \@firstoftwo
 \else \expandafter \@secondoftwo
 \fi
}%
\providecommand \natexlab [1]{#1}%
\providecommand \enquote  [1]{``#1''}%
\providecommand \bibnamefont  [1]{#1}%
\providecommand \bibfnamefont [1]{#1}%
\providecommand \citenamefont [1]{#1}%
\providecommand \href@noop [0]{\@secondoftwo}%
\providecommand \href [0]{\begingroup \@sanitize@url \@href}%
\providecommand \@href[1]{\@@startlink{#1}\@@href}%
\providecommand \@@href[1]{\endgroup#1\@@endlink}%
\providecommand \@sanitize@url [0]{\catcode `\\12\catcode `\$12\catcode
  `\&12\catcode `\#12\catcode `\^12\catcode `\_12\catcode `\%12\relax}%
\providecommand \@@startlink[1]{}%
\providecommand \@@endlink[0]{}%
\providecommand \url  [0]{\begingroup\@sanitize@url \@url }%
\providecommand \@url [1]{\endgroup\@href {#1}{\urlprefix }}%
\providecommand \urlprefix  [0]{URL }%
\providecommand \Eprint [0]{\href }%
\providecommand \doibase [0]{https://doi.org/}%
\providecommand \selectlanguage [0]{\@gobble}%
\providecommand \bibinfo  [0]{\@secondoftwo}%
\providecommand \bibfield  [0]{\@secondoftwo}%
\providecommand \translation [1]{[#1]}%
\providecommand \BibitemOpen [0]{}%
\providecommand \bibitemStop [0]{}%
\providecommand \bibitemNoStop [0]{.\EOS\space}%
\providecommand \EOS [0]{\spacefactor3000\relax}%
\providecommand \BibitemShut  [1]{\csname bibitem#1\endcsname}%
\let\auto@bib@innerbib\@empty
\bibitem [{\citenamefont {Arute}\ \emph {et~al.}(2019)\citenamefont {Arute},
  \citenamefont {Arya}, \citenamefont {Babbush}, \citenamefont {Bacon},
  \citenamefont {Bardin}, \citenamefont {Barends}, \citenamefont {Biswas},
  \citenamefont {Boixo}, \citenamefont {Brandao}, \citenamefont {Buell} \emph
  {et~al.}}]{supermacy_Martinis}%
  \BibitemOpen
  \bibfield  {author} {\bibinfo {author} {\bibfnamefont {F.}~\bibnamefont
  {Arute}}, \bibinfo {author} {\bibfnamefont {K.}~\bibnamefont {Arya}},
  \bibinfo {author} {\bibfnamefont {R.}~\bibnamefont {Babbush}}, \bibinfo
  {author} {\bibfnamefont {D.}~\bibnamefont {Bacon}}, \bibinfo {author}
  {\bibfnamefont {J.~C.}\ \bibnamefont {Bardin}}, \bibinfo {author}
  {\bibfnamefont {R.}~\bibnamefont {Barends}}, \bibinfo {author} {\bibfnamefont
  {R.}~\bibnamefont {Biswas}}, \bibinfo {author} {\bibfnamefont
  {S.}~\bibnamefont {Boixo}}, \bibinfo {author} {\bibfnamefont {F.~G.}\
  \bibnamefont {Brandao}}, \bibinfo {author} {\bibfnamefont {D.~A.}\
  \bibnamefont {Buell}}, \emph {et~al.},\ }\bibfield  {title} {\bibinfo {title}
  {Quantum supremacy using a programmable superconducting processor},\
  }\href@noop {} {\bibfield  {journal} {\bibinfo  {journal} {Nature}\ }\textbf
  {\bibinfo {volume} {574}},\ \bibinfo {pages} {505} (\bibinfo {year}
  {2019})}\BibitemShut {NoStop}%
\bibitem [{\citenamefont {Wu}\ \emph {et~al.}(2021)\citenamefont {Wu},
  \citenamefont {Bao}, \citenamefont {Cao}, \citenamefont {Chen}, \citenamefont
  {Chen}, \citenamefont {Chen}, \citenamefont {Chung}, \citenamefont {Deng},
  \citenamefont {Du}, \citenamefont {Fan} \emph {et~al.}}]{wu2021strong}%
  \BibitemOpen
  \bibfield  {author} {\bibinfo {author} {\bibfnamefont {Y.}~\bibnamefont
  {Wu}}, \bibinfo {author} {\bibfnamefont {W.-S.}\ \bibnamefont {Bao}},
  \bibinfo {author} {\bibfnamefont {S.}~\bibnamefont {Cao}}, \bibinfo {author}
  {\bibfnamefont {F.}~\bibnamefont {Chen}}, \bibinfo {author} {\bibfnamefont
  {M.-C.}\ \bibnamefont {Chen}}, \bibinfo {author} {\bibfnamefont
  {X.}~\bibnamefont {Chen}}, \bibinfo {author} {\bibfnamefont {T.-H.}\
  \bibnamefont {Chung}}, \bibinfo {author} {\bibfnamefont {H.}~\bibnamefont
  {Deng}}, \bibinfo {author} {\bibfnamefont {Y.}~\bibnamefont {Du}}, \bibinfo
  {author} {\bibfnamefont {D.}~\bibnamefont {Fan}}, \emph {et~al.},\ }\bibfield
   {title} {\bibinfo {title} {Strong quantum computational advantage using a
  superconducting quantum processor},\ }\href@noop {} {\bibfield  {journal}
  {\bibinfo  {journal} {Physical review letters}\ }\textbf {\bibinfo {volume}
  {127}},\ \bibinfo {pages} {180501} (\bibinfo {year} {2021})}\BibitemShut
  {NoStop}%
\bibitem [{\citenamefont {AI}(2023)}]{surface_google}%
  \BibitemOpen
  \bibfield  {author} {\bibinfo {author} {\bibfnamefont {G.~Q.}\ \bibnamefont
  {AI}},\ }\bibfield  {title} {\bibinfo {title} {Suppressing quantum errors by
  scaling a surface code logical qubit},\ }\href@noop {} {\bibfield  {journal}
  {\bibinfo  {journal} {Nature}\ }\textbf {\bibinfo {volume} {614}},\ \bibinfo
  {pages} {676} (\bibinfo {year} {2023})}\BibitemShut {NoStop}%
\bibitem [{\citenamefont {Kim}\ \emph {et~al.}(2023)\citenamefont {Kim},
  \citenamefont {Eddins}, \citenamefont {Anand}, \citenamefont {Wei},
  \citenamefont {Van Den~Berg}, \citenamefont {Rosenblatt}, \citenamefont
  {Nayfeh}, \citenamefont {Wu}, \citenamefont {Zaletel}, \citenamefont {Temme}
  \emph {et~al.}}]{kim2023evidence}%
  \BibitemOpen
  \bibfield  {author} {\bibinfo {author} {\bibfnamefont {Y.}~\bibnamefont
  {Kim}}, \bibinfo {author} {\bibfnamefont {A.}~\bibnamefont {Eddins}},
  \bibinfo {author} {\bibfnamefont {S.}~\bibnamefont {Anand}}, \bibinfo
  {author} {\bibfnamefont {K.~X.}\ \bibnamefont {Wei}}, \bibinfo {author}
  {\bibfnamefont {E.}~\bibnamefont {Van Den~Berg}}, \bibinfo {author}
  {\bibfnamefont {S.}~\bibnamefont {Rosenblatt}}, \bibinfo {author}
  {\bibfnamefont {H.}~\bibnamefont {Nayfeh}}, \bibinfo {author} {\bibfnamefont
  {Y.}~\bibnamefont {Wu}}, \bibinfo {author} {\bibfnamefont {M.}~\bibnamefont
  {Zaletel}}, \bibinfo {author} {\bibfnamefont {K.}~\bibnamefont {Temme}},
  \emph {et~al.},\ }\bibfield  {title} {\bibinfo {title} {Evidence for the
  utility of quantum computing before fault tolerance},\ }\href@noop {}
  {\bibfield  {journal} {\bibinfo  {journal} {Nature}\ }\textbf {\bibinfo
  {volume} {618}},\ \bibinfo {pages} {500} (\bibinfo {year}
  {2023})}\BibitemShut {NoStop}%
\bibitem [{\citenamefont {Rosen}\ \emph {et~al.}()\citenamefont {Rosen},
  \citenamefont {Muschinske}, \citenamefont {Barrett}, \citenamefont
  {Chatterjee}, \citenamefont {Hays}, \citenamefont {DeMarco}, \citenamefont
  {Karamlou}, \citenamefont {Rower}, \citenamefont {Das}, \citenamefont {Kim},
  \citenamefont {Niedzielski}, \citenamefont {Schuldt}, \citenamefont
  {Serniak}, \citenamefont {Schwartz}, \citenamefont {Yoder}, \citenamefont
  {Grover},\ and\ \citenamefont
  {Oliver}}]{rosen2024implementingsyntheticmagneticvector}%
  \BibitemOpen
  \bibfield  {author} {\bibinfo {author} {\bibfnamefont {I.~T.}\ \bibnamefont
  {Rosen}}, \bibinfo {author} {\bibfnamefont {S.}~\bibnamefont {Muschinske}},
  \bibinfo {author} {\bibfnamefont {C.~N.}\ \bibnamefont {Barrett}}, \bibinfo
  {author} {\bibfnamefont {A.}~\bibnamefont {Chatterjee}}, \bibinfo {author}
  {\bibfnamefont {M.}~\bibnamefont {Hays}}, \bibinfo {author} {\bibfnamefont
  {M.}~\bibnamefont {DeMarco}}, \bibinfo {author} {\bibfnamefont
  {A.}~\bibnamefont {Karamlou}}, \bibinfo {author} {\bibfnamefont
  {D.}~\bibnamefont {Rower}}, \bibinfo {author} {\bibfnamefont
  {R.}~\bibnamefont {Das}}, \bibinfo {author} {\bibfnamefont {D.~K.}\
  \bibnamefont {Kim}}, \bibinfo {author} {\bibfnamefont {B.~M.}\ \bibnamefont
  {Niedzielski}}, \bibinfo {author} {\bibfnamefont {M.}~\bibnamefont
  {Schuldt}}, \bibinfo {author} {\bibfnamefont {K.}~\bibnamefont {Serniak}},
  \bibinfo {author} {\bibfnamefont {M.~E.}\ \bibnamefont {Schwartz}}, \bibinfo
  {author} {\bibfnamefont {J.~L.}\ \bibnamefont {Yoder}}, \bibinfo {author}
  {\bibfnamefont {J.~A.}\ \bibnamefont {Grover}},\ and\ \bibinfo {author}
  {\bibfnamefont {W.~D.}\ \bibnamefont {Oliver}},\ }\bibfield  {title}
  {\bibinfo {title} {Implementing a synthetic magnetic vector potential in a 2d
  superconducting qubit array},\ }\href@noop {} {\bibinfo  {journal} {Preprint
  at https://arxiv.org/abs/2405.00873}\ }\BibitemShut {NoStop}%
\bibitem [{\citenamefont {Fowler}\ \emph {et~al.}(2012)\citenamefont {Fowler},
  \citenamefont {Mariantoni}, \citenamefont {Martinis},\ and\ \citenamefont
  {Cleland}}]{surface_theo}%
  \BibitemOpen
\bibfield  {journal} {  }\bibfield  {author} {\bibinfo {author} {\bibfnamefont
  {A.~G.}\ \bibnamefont {Fowler}}, \bibinfo {author} {\bibfnamefont
  {M.}~\bibnamefont {Mariantoni}}, \bibinfo {author} {\bibfnamefont {J.~M.}\
  \bibnamefont {Martinis}},\ and\ \bibinfo {author} {\bibfnamefont {A.~N.}\
  \bibnamefont {Cleland}},\ }\bibfield  {title} {\bibinfo {title} {Surface
  codes: Towards practical large-scale quantum computation},\ }\href@noop {}
  {\bibfield  {journal} {\bibinfo  {journal} {Phys. Rev. A}\ }\textbf {\bibinfo
  {volume} {86}},\ \bibinfo {pages} {032324} (\bibinfo {year}
  {2012})}\BibitemShut {NoStop}%
\bibitem [{\citenamefont {Bravyi}\ \emph {et~al.}(2024)\citenamefont {Bravyi},
  \citenamefont {Cross}, \citenamefont {Gambetta}, \citenamefont {Maslov},
  \citenamefont {Rall},\ and\ \citenamefont {Yoder}}]{QEC}%
  \BibitemOpen
  \bibfield  {author} {\bibinfo {author} {\bibfnamefont {S.}~\bibnamefont
  {Bravyi}}, \bibinfo {author} {\bibfnamefont {A.~W.}\ \bibnamefont {Cross}},
  \bibinfo {author} {\bibfnamefont {J.~M.}\ \bibnamefont {Gambetta}}, \bibinfo
  {author} {\bibfnamefont {D.}~\bibnamefont {Maslov}}, \bibinfo {author}
  {\bibfnamefont {P.}~\bibnamefont {Rall}},\ and\ \bibinfo {author}
  {\bibfnamefont {T.~J.}\ \bibnamefont {Yoder}},\ }\bibfield  {title} {\bibinfo
  {title} {High-threshold and low-overhead fault-tolerant quantum memory},\
  }\href@noop {} {\bibfield  {journal} {\bibinfo  {journal} {Nature}\ }\textbf
  {\bibinfo {volume} {627}},\ \bibinfo {pages} {778} (\bibinfo {year}
  {2024})}\BibitemShut {NoStop}%
\bibitem [{\citenamefont {Jiang}\ \emph {et~al.}(2007)\citenamefont {Jiang},
  \citenamefont {Taylor}, \citenamefont {S{\o}rensen},\ and\ \citenamefont
  {Lukin}}]{distributed_Lukin}%
  \BibitemOpen
  \bibfield  {author} {\bibinfo {author} {\bibfnamefont {L.}~\bibnamefont
  {Jiang}}, \bibinfo {author} {\bibfnamefont {J.~M.}\ \bibnamefont {Taylor}},
  \bibinfo {author} {\bibfnamefont {A.~S.}\ \bibnamefont {S{\o}rensen}},\ and\
  \bibinfo {author} {\bibfnamefont {M.~D.}\ \bibnamefont {Lukin}},\ }\bibfield
  {title} {\bibinfo {title} {Distributed quantum computation based on small
  quantum registers},\ }\href@noop {} {\bibfield  {journal} {\bibinfo
  {journal} {Phys. Rev. A}\ }\textbf {\bibinfo {volume} {76}},\ \bibinfo
  {pages} {062323} (\bibinfo {year} {2007})}\BibitemShut {NoStop}%
\bibitem [{\citenamefont {Monroe}\ \emph {et~al.}(2014)\citenamefont {Monroe},
  \citenamefont {Raussendorf}, \citenamefont {Ruthven}, \citenamefont {Brown},
  \citenamefont {Maunz}, \citenamefont {Duan},\ and\ \citenamefont
  {Kim}}]{largescale_luming}%
  \BibitemOpen
  \bibfield  {author} {\bibinfo {author} {\bibfnamefont {C.}~\bibnamefont
  {Monroe}}, \bibinfo {author} {\bibfnamefont {R.}~\bibnamefont {Raussendorf}},
  \bibinfo {author} {\bibfnamefont {A.}~\bibnamefont {Ruthven}}, \bibinfo
  {author} {\bibfnamefont {K.~R.}\ \bibnamefont {Brown}}, \bibinfo {author}
  {\bibfnamefont {P.}~\bibnamefont {Maunz}}, \bibinfo {author} {\bibfnamefont
  {L.-M.}\ \bibnamefont {Duan}},\ and\ \bibinfo {author} {\bibfnamefont
  {J.}~\bibnamefont {Kim}},\ }\bibfield  {title} {\bibinfo {title} {Large-scale
  modular quantum-computer architecture with atomic memory and photonic
  interconnects},\ }\href@noop {} {\bibfield  {journal} {\bibinfo  {journal}
  {Phys. Rev. A}\ }\textbf {\bibinfo {volume} {89}},\ \bibinfo {pages} {022317}
  (\bibinfo {year} {2014})}\BibitemShut {NoStop}%
\bibitem [{\citenamefont {Gold}\ \emph {et~al.}(2021)\citenamefont {Gold},
  \citenamefont {Paquette}, \citenamefont {Stockklauser}, \citenamefont
  {Reagor}, \citenamefont {Alam}, \citenamefont {Bestwick}, \citenamefont
  {Didier}, \citenamefont {Nersisyan}, \citenamefont {Oruc}, \citenamefont
  {Razavi} \emph {et~al.}}]{silicondie_rigetti}%
  \BibitemOpen
  \bibfield  {author} {\bibinfo {author} {\bibfnamefont {A.}~\bibnamefont
  {Gold}}, \bibinfo {author} {\bibfnamefont {J.}~\bibnamefont {Paquette}},
  \bibinfo {author} {\bibfnamefont {A.}~\bibnamefont {Stockklauser}}, \bibinfo
  {author} {\bibfnamefont {M.~J.}\ \bibnamefont {Reagor}}, \bibinfo {author}
  {\bibfnamefont {M.~S.}\ \bibnamefont {Alam}}, \bibinfo {author}
  {\bibfnamefont {A.}~\bibnamefont {Bestwick}}, \bibinfo {author}
  {\bibfnamefont {N.}~\bibnamefont {Didier}}, \bibinfo {author} {\bibfnamefont
  {A.}~\bibnamefont {Nersisyan}}, \bibinfo {author} {\bibfnamefont
  {F.}~\bibnamefont {Oruc}}, \bibinfo {author} {\bibfnamefont {A.}~\bibnamefont
  {Razavi}}, \emph {et~al.},\ }\bibfield  {title} {\bibinfo {title}
  {Entanglement across separate silicon dies in a modular superconducting qubit
  device},\ }\href@noop {} {\bibfield  {journal} {\bibinfo  {journal} {npj
  Quantum Inf.}\ }\textbf {\bibinfo {volume} {7}},\ \bibinfo {pages} {142}
  (\bibinfo {year} {2021})}\BibitemShut {NoStop}%
\bibitem [{\citenamefont {Ang}\ \emph {et~al.}()\citenamefont {Ang},
  \citenamefont {Carini}, \citenamefont {Chen}, \citenamefont {Chuang},
  \citenamefont {DeMarco}, \citenamefont {Economou}, \citenamefont {Eickbusch},
  \citenamefont {Faraon}, \citenamefont {Fu}, \citenamefont {Girvin} \emph
  {et~al.}}]{ARQUIN_architectures}%
  \BibitemOpen
  \bibfield  {author} {\bibinfo {author} {\bibfnamefont {J.}~\bibnamefont
  {Ang}}, \bibinfo {author} {\bibfnamefont {G.}~\bibnamefont {Carini}},
  \bibinfo {author} {\bibfnamefont {Y.}~\bibnamefont {Chen}}, \bibinfo {author}
  {\bibfnamefont {I.}~\bibnamefont {Chuang}}, \bibinfo {author} {\bibfnamefont
  {M.}~\bibnamefont {DeMarco}}, \bibinfo {author} {\bibfnamefont
  {S.}~\bibnamefont {Economou}}, \bibinfo {author} {\bibfnamefont
  {A.}~\bibnamefont {Eickbusch}}, \bibinfo {author} {\bibfnamefont
  {A.}~\bibnamefont {Faraon}}, \bibinfo {author} {\bibfnamefont {K.-M.}\
  \bibnamefont {Fu}}, \bibinfo {author} {\bibfnamefont {S.}~\bibnamefont
  {Girvin}}, \emph {et~al.},\ }\bibfield  {title} {\bibinfo {title} {Arquin:
  architectures for multinode superconducting quantum computers},\ }\href@noop
  {} {\bibinfo  {journal} {ACM Trans. Quantum Comput.}\ }\BibitemShut {NoStop}%
\bibitem [{\citenamefont {Barenco}\ \emph {et~al.}(1995)\citenamefont
  {Barenco}, \citenamefont {Bennett}, \citenamefont {Cleve}, \citenamefont
  {DiVincenzo}, \citenamefont {Margolus}, \citenamefont {Shor}, \citenamefont
  {Sleator}, \citenamefont {Smolin},\ and\ \citenamefont
  {Weinfurter}}]{barenco1995elementary}%
  \BibitemOpen
\bibfield  {journal} {  }\bibfield  {author} {\bibinfo {author} {\bibfnamefont
  {A.}~\bibnamefont {Barenco}}, \bibinfo {author} {\bibfnamefont {C.~H.}\
  \bibnamefont {Bennett}}, \bibinfo {author} {\bibfnamefont {R.}~\bibnamefont
  {Cleve}}, \bibinfo {author} {\bibfnamefont {D.~P.}\ \bibnamefont
  {DiVincenzo}}, \bibinfo {author} {\bibfnamefont {N.}~\bibnamefont
  {Margolus}}, \bibinfo {author} {\bibfnamefont {P.}~\bibnamefont {Shor}},
  \bibinfo {author} {\bibfnamefont {T.}~\bibnamefont {Sleator}}, \bibinfo
  {author} {\bibfnamefont {J.~A.}\ \bibnamefont {Smolin}},\ and\ \bibinfo
  {author} {\bibfnamefont {H.}~\bibnamefont {Weinfurter}},\ }\bibfield  {title}
  {\bibinfo {title} {Elementary gates for quantum computation},\ }\href
  {https://doi.org/10.1103/PhysRevA.52.3457} {\bibfield  {journal} {\bibinfo
  {journal} {Phys. Rev. A}\ }\textbf {\bibinfo {volume} {52}},\ \bibinfo
  {pages} {3457} (\bibinfo {year} {1995})}\BibitemShut {NoStop}%
\bibitem [{\citenamefont {Leung}\ \emph {et~al.}(2019)\citenamefont {Leung},
  \citenamefont {Lu}, \citenamefont {Chakram}, \citenamefont {Naik},
  \citenamefont {Earnest}, \citenamefont {Ma}, \citenamefont {Jacobs},
  \citenamefont {Cleland},\ and\ \citenamefont
  {Schuster}}]{bidiractional_cleland}%
  \BibitemOpen
  \bibfield  {author} {\bibinfo {author} {\bibfnamefont {N.}~\bibnamefont
  {Leung}}, \bibinfo {author} {\bibfnamefont {Y.}~\bibnamefont {Lu}}, \bibinfo
  {author} {\bibfnamefont {S.}~\bibnamefont {Chakram}}, \bibinfo {author}
  {\bibfnamefont {R.}~\bibnamefont {Naik}}, \bibinfo {author} {\bibfnamefont
  {N.}~\bibnamefont {Earnest}}, \bibinfo {author} {\bibfnamefont
  {R.}~\bibnamefont {Ma}}, \bibinfo {author} {\bibfnamefont {K.}~\bibnamefont
  {Jacobs}}, \bibinfo {author} {\bibfnamefont {A.}~\bibnamefont {Cleland}},\
  and\ \bibinfo {author} {\bibfnamefont {D.}~\bibnamefont {Schuster}},\
  }\bibfield  {title} {\bibinfo {title} {Deterministic bidirectional
  communication and remote entanglement generation between superconducting
  qubits},\ }\href@noop {} {\bibfield  {journal} {\bibinfo  {journal} {npj
  Quantum Inf.}\ }\textbf {\bibinfo {volume} {5}},\ \bibinfo {pages} {18}
  (\bibinfo {year} {2019})}\BibitemShut {NoStop}%
\bibitem [{\citenamefont {Zhong}\ \emph {et~al.}(2019)\citenamefont {Zhong},
  \citenamefont {Chang}, \citenamefont {Satzinger}, \citenamefont {Chou},
  \citenamefont {Bienfait}, \citenamefont {Conner}, \citenamefont {Dumur},
  \citenamefont {Grebel}, \citenamefont {Peairs}, \citenamefont {Povey} \emph
  {et~al.}}]{violateBell_zhong}%
  \BibitemOpen
  \bibfield  {author} {\bibinfo {author} {\bibfnamefont {Y.}~\bibnamefont
  {Zhong}}, \bibinfo {author} {\bibfnamefont {H.-S.}\ \bibnamefont {Chang}},
  \bibinfo {author} {\bibfnamefont {K.}~\bibnamefont {Satzinger}}, \bibinfo
  {author} {\bibfnamefont {M.-H.}\ \bibnamefont {Chou}}, \bibinfo {author}
  {\bibfnamefont {A.}~\bibnamefont {Bienfait}}, \bibinfo {author}
  {\bibfnamefont {C.}~\bibnamefont {Conner}}, \bibinfo {author} {\bibfnamefont
  {{\'E}.}~\bibnamefont {Dumur}}, \bibinfo {author} {\bibfnamefont
  {J.}~\bibnamefont {Grebel}}, \bibinfo {author} {\bibfnamefont
  {G.}~\bibnamefont {Peairs}}, \bibinfo {author} {\bibfnamefont
  {R.}~\bibnamefont {Povey}}, \emph {et~al.},\ }\bibfield  {title} {\bibinfo
  {title} {Violating bell’s inequality with remotely connected
  superconducting qubits},\ }\href@noop {} {\bibfield  {journal} {\bibinfo
  {journal} {Nat. Phys.}\ }\textbf {\bibinfo {volume} {15}},\ \bibinfo {pages}
  {741} (\bibinfo {year} {2019})}\BibitemShut {NoStop}%
\bibitem [{\citenamefont {Magnard}\ \emph {et~al.}(2020)\citenamefont
  {Magnard}, \citenamefont {Storz}, \citenamefont {Kurpiers}, \citenamefont
  {Sch{\"a}r}, \citenamefont {Marxer}, \citenamefont {L{\"u}tolf},
  \citenamefont {Walter}, \citenamefont {Besse}, \citenamefont {Gabureac},
  \citenamefont {Reuer} \emph {et~al.}}]{quantumlink_walraff}%
  \BibitemOpen
  \bibfield  {author} {\bibinfo {author} {\bibfnamefont {P.}~\bibnamefont
  {Magnard}}, \bibinfo {author} {\bibfnamefont {S.}~\bibnamefont {Storz}},
  \bibinfo {author} {\bibfnamefont {P.}~\bibnamefont {Kurpiers}}, \bibinfo
  {author} {\bibfnamefont {J.}~\bibnamefont {Sch{\"a}r}}, \bibinfo {author}
  {\bibfnamefont {F.}~\bibnamefont {Marxer}}, \bibinfo {author} {\bibfnamefont
  {J.}~\bibnamefont {L{\"u}tolf}}, \bibinfo {author} {\bibfnamefont
  {T.}~\bibnamefont {Walter}}, \bibinfo {author} {\bibfnamefont {J.-C.}\
  \bibnamefont {Besse}}, \bibinfo {author} {\bibfnamefont {M.}~\bibnamefont
  {Gabureac}}, \bibinfo {author} {\bibfnamefont {K.}~\bibnamefont {Reuer}},
  \emph {et~al.},\ }\bibfield  {title} {\bibinfo {title} {Microwave quantum
  link between superconducting circuits housed in spatially separated cryogenic
  systems},\ }\href@noop {} {\bibfield  {journal} {\bibinfo  {journal} {Phys.
  Rev. Lett.}\ }\textbf {\bibinfo {volume} {125}},\ \bibinfo {pages} {260502}
  (\bibinfo {year} {2020})}\BibitemShut {NoStop}%
\bibitem [{\citenamefont {Campagne-Ibarcq}\ \emph {et~al.}(2018)\citenamefont
  {Campagne-Ibarcq}, \citenamefont {Zalys-Geller}, \citenamefont {Narla},
  \citenamefont {Shankar}, \citenamefont {Reinhold}, \citenamefont {Burkhart},
  \citenamefont {Axline}, \citenamefont {Pfaff}, \citenamefont {Frunzio},
  \citenamefont {Schoelkopf},\ and\ \citenamefont
  {Devoret}}]{Devoret_deterministic}%
  \BibitemOpen
  \bibfield  {author} {\bibinfo {author} {\bibfnamefont {P.}~\bibnamefont
  {Campagne-Ibarcq}}, \bibinfo {author} {\bibfnamefont {E.}~\bibnamefont
  {Zalys-Geller}}, \bibinfo {author} {\bibfnamefont {A.}~\bibnamefont {Narla}},
  \bibinfo {author} {\bibfnamefont {S.}~\bibnamefont {Shankar}}, \bibinfo
  {author} {\bibfnamefont {P.}~\bibnamefont {Reinhold}}, \bibinfo {author}
  {\bibfnamefont {L.}~\bibnamefont {Burkhart}}, \bibinfo {author}
  {\bibfnamefont {C.}~\bibnamefont {Axline}}, \bibinfo {author} {\bibfnamefont
  {W.}~\bibnamefont {Pfaff}}, \bibinfo {author} {\bibfnamefont
  {L.}~\bibnamefont {Frunzio}}, \bibinfo {author} {\bibfnamefont {R.~J.}\
  \bibnamefont {Schoelkopf}},\ and\ \bibinfo {author} {\bibfnamefont {M.~H.}\
  \bibnamefont {Devoret}},\ }\bibfield  {title} {\bibinfo {title}
  {Deterministic remote entanglement of superconducting circuits through
  microwave two-photon transitions},\ }\href@noop {} {\bibfield  {journal}
  {\bibinfo  {journal} {Phys. Rev. Lett.}\ }\textbf {\bibinfo {volume} {120}},\
  \bibinfo {pages} {200501} (\bibinfo {year} {2018})}\BibitemShut {NoStop}%
\bibitem [{\citenamefont {Zhong}\ \emph {et~al.}(2021)\citenamefont {Zhong},
  \citenamefont {Chang}, \citenamefont {Bienfait}, \citenamefont {Dumur},
  \citenamefont {Chou}, \citenamefont {Conner}, \citenamefont {Grebel},
  \citenamefont {Povey}, \citenamefont {Yan}, \citenamefont {Schuster} \emph
  {et~al.}}]{deterministicentanglement_zhong}%
  \BibitemOpen
  \bibfield  {author} {\bibinfo {author} {\bibfnamefont {Y.}~\bibnamefont
  {Zhong}}, \bibinfo {author} {\bibfnamefont {H.-S.}\ \bibnamefont {Chang}},
  \bibinfo {author} {\bibfnamefont {A.}~\bibnamefont {Bienfait}}, \bibinfo
  {author} {\bibfnamefont {{\'E}.}~\bibnamefont {Dumur}}, \bibinfo {author}
  {\bibfnamefont {M.-H.}\ \bibnamefont {Chou}}, \bibinfo {author}
  {\bibfnamefont {C.~R.}\ \bibnamefont {Conner}}, \bibinfo {author}
  {\bibfnamefont {J.}~\bibnamefont {Grebel}}, \bibinfo {author} {\bibfnamefont
  {R.~G.}\ \bibnamefont {Povey}}, \bibinfo {author} {\bibfnamefont
  {H.}~\bibnamefont {Yan}}, \bibinfo {author} {\bibfnamefont {D.~I.}\
  \bibnamefont {Schuster}}, \emph {et~al.},\ }\bibfield  {title} {\bibinfo
  {title} {Deterministic multi-qubit entanglement in a quantum network},\
  }\href@noop {} {\bibfield  {journal} {\bibinfo  {journal} {Nature}\ }\textbf
  {\bibinfo {volume} {590}},\ \bibinfo {pages} {571} (\bibinfo {year}
  {2021})}\BibitemShut {NoStop}%
\bibitem [{\citenamefont {Yan}\ \emph {et~al.}(2022)\citenamefont {Yan},
  \citenamefont {Zhong}, \citenamefont {Chang}, \citenamefont {Bienfait},
  \citenamefont {Chou}, \citenamefont {Conner}, \citenamefont {Dumur},
  \citenamefont {Grebel}, \citenamefont {Povey},\ and\ \citenamefont
  {Cleland}}]{purification_zhong}%
  \BibitemOpen
  \bibfield  {author} {\bibinfo {author} {\bibfnamefont {H.}~\bibnamefont
  {Yan}}, \bibinfo {author} {\bibfnamefont {Y.}~\bibnamefont {Zhong}}, \bibinfo
  {author} {\bibfnamefont {H.-S.}\ \bibnamefont {Chang}}, \bibinfo {author}
  {\bibfnamefont {A.}~\bibnamefont {Bienfait}}, \bibinfo {author}
  {\bibfnamefont {M.-H.}\ \bibnamefont {Chou}}, \bibinfo {author}
  {\bibfnamefont {C.~R.}\ \bibnamefont {Conner}}, \bibinfo {author}
  {\bibfnamefont {{\'E}.}~\bibnamefont {Dumur}}, \bibinfo {author}
  {\bibfnamefont {J.}~\bibnamefont {Grebel}}, \bibinfo {author} {\bibfnamefont
  {R.~G.}\ \bibnamefont {Povey}},\ and\ \bibinfo {author} {\bibfnamefont
  {A.~N.}\ \bibnamefont {Cleland}},\ }\bibfield  {title} {\bibinfo {title}
  {Entanglement purification and protection in a superconducting quantum
  network},\ }\href@noop {} {\bibfield  {journal} {\bibinfo  {journal} {Phys.
  Rev. Lett.}\ }\textbf {\bibinfo {volume} {128}},\ \bibinfo {pages} {080504}
  (\bibinfo {year} {2022})}\BibitemShut {NoStop}%
\bibitem [{\citenamefont {Storz}\ \emph {et~al.}(2023)\citenamefont {Storz},
  \citenamefont {Sch{\"a}r}, \citenamefont {Kulikov}, \citenamefont {Magnard},
  \citenamefont {Kurpiers}, \citenamefont {L{\"u}tolf}, \citenamefont {Walter},
  \citenamefont {Copetudo}, \citenamefont {Reuer}, \citenamefont {Akin} \emph
  {et~al.}}]{loopholefree_walraff}%
  \BibitemOpen
  \bibfield  {author} {\bibinfo {author} {\bibfnamefont {S.}~\bibnamefont
  {Storz}}, \bibinfo {author} {\bibfnamefont {J.}~\bibnamefont {Sch{\"a}r}},
  \bibinfo {author} {\bibfnamefont {A.}~\bibnamefont {Kulikov}}, \bibinfo
  {author} {\bibfnamefont {P.}~\bibnamefont {Magnard}}, \bibinfo {author}
  {\bibfnamefont {P.}~\bibnamefont {Kurpiers}}, \bibinfo {author}
  {\bibfnamefont {J.}~\bibnamefont {L{\"u}tolf}}, \bibinfo {author}
  {\bibfnamefont {T.}~\bibnamefont {Walter}}, \bibinfo {author} {\bibfnamefont
  {A.}~\bibnamefont {Copetudo}}, \bibinfo {author} {\bibfnamefont
  {K.}~\bibnamefont {Reuer}}, \bibinfo {author} {\bibfnamefont
  {A.}~\bibnamefont {Akin}}, \emph {et~al.},\ }\bibfield  {title} {\bibinfo
  {title} {Loophole-free bell inequality violation with superconducting
  circuits},\ }\href@noop {} {\bibfield  {journal} {\bibinfo  {journal}
  {Nature}\ }\textbf {\bibinfo {volume} {617}},\ \bibinfo {pages} {265}
  (\bibinfo {year} {2023})}\BibitemShut {NoStop}%
\bibitem [{\citenamefont {Chang}\ \emph {et~al.}(2020)\citenamefont {Chang},
  \citenamefont {Zhong}, \citenamefont {Bienfait}, \citenamefont {Chou},
  \citenamefont {Conner}, \citenamefont {Dumur}, \citenamefont {Grebel},
  \citenamefont {Peairs}, \citenamefont {Povey}, \citenamefont {Satzinger},\
  and\ \citenamefont {Cleland}}]{remote_Chang}%
  \BibitemOpen
  \bibfield  {author} {\bibinfo {author} {\bibfnamefont {H.-S.}\ \bibnamefont
  {Chang}}, \bibinfo {author} {\bibfnamefont {Y.~P.}\ \bibnamefont {Zhong}},
  \bibinfo {author} {\bibfnamefont {A.}~\bibnamefont {Bienfait}}, \bibinfo
  {author} {\bibfnamefont {M.-H.}\ \bibnamefont {Chou}}, \bibinfo {author}
  {\bibfnamefont {C.~R.}\ \bibnamefont {Conner}}, \bibinfo {author}
  {\bibfnamefont {E.}~\bibnamefont {Dumur}}, \bibinfo {author} {\bibfnamefont
  {J.}~\bibnamefont {Grebel}}, \bibinfo {author} {\bibfnamefont {G.~A.}\
  \bibnamefont {Peairs}}, \bibinfo {author} {\bibfnamefont {R.~G.}\
  \bibnamefont {Povey}}, \bibinfo {author} {\bibfnamefont {K.~J.}\ \bibnamefont
  {Satzinger}},\ and\ \bibinfo {author} {\bibfnamefont {A.~N.}\ \bibnamefont
  {Cleland}},\ }\bibfield  {title} {\bibinfo {title} {Remote entanglement via
  adiabatic passage using a tunably dissipative quantum communication system},\
  }\href@noop {} {\bibfield  {journal} {\bibinfo  {journal} {Phys. Rev. Lett.}\
  }\textbf {\bibinfo {volume} {124}},\ \bibinfo {pages} {240502} (\bibinfo
  {year} {2020})}\BibitemShut {NoStop}%
\bibitem [{\citenamefont {Niu}\ \emph {et~al.}(2023)\citenamefont {Niu},
  \citenamefont {Zhang}, \citenamefont {Liu}, \citenamefont {Qiu},
  \citenamefont {Huang}, \citenamefont {Huang}, \citenamefont {Jia},
  \citenamefont {Liu}, \citenamefont {Tao}, \citenamefont {Wei} \emph
  {et~al.}}]{lowloss_zhong}%
  \BibitemOpen
  \bibfield  {author} {\bibinfo {author} {\bibfnamefont {J.}~\bibnamefont
  {Niu}}, \bibinfo {author} {\bibfnamefont {L.}~\bibnamefont {Zhang}}, \bibinfo
  {author} {\bibfnamefont {Y.}~\bibnamefont {Liu}}, \bibinfo {author}
  {\bibfnamefont {J.}~\bibnamefont {Qiu}}, \bibinfo {author} {\bibfnamefont
  {W.}~\bibnamefont {Huang}}, \bibinfo {author} {\bibfnamefont
  {J.}~\bibnamefont {Huang}}, \bibinfo {author} {\bibfnamefont
  {H.}~\bibnamefont {Jia}}, \bibinfo {author} {\bibfnamefont {J.}~\bibnamefont
  {Liu}}, \bibinfo {author} {\bibfnamefont {Z.}~\bibnamefont {Tao}}, \bibinfo
  {author} {\bibfnamefont {W.}~\bibnamefont {Wei}}, \emph {et~al.},\ }\bibfield
   {title} {\bibinfo {title} {Low-loss interconnects for modular
  superconducting quantum processors},\ }\href@noop {} {\bibfield  {journal}
  {\bibinfo  {journal} {Nat. Electron.}\ }\textbf {\bibinfo {volume} {6}},\
  \bibinfo {pages} {235} (\bibinfo {year} {2023})}\BibitemShut {NoStop}%
\bibitem [{\citenamefont {Zhang}\ \emph {et~al.}(2003)\citenamefont {Zhang},
  \citenamefont {Vala}, \citenamefont {Sastry},\ and\ \citenamefont
  {Whaley}}]{perfectentangler_zhang}%
  \BibitemOpen
  \bibfield  {author} {\bibinfo {author} {\bibfnamefont {J.}~\bibnamefont
  {Zhang}}, \bibinfo {author} {\bibfnamefont {J.}~\bibnamefont {Vala}},
  \bibinfo {author} {\bibfnamefont {S.}~\bibnamefont {Sastry}},\ and\ \bibinfo
  {author} {\bibfnamefont {K.~B.}\ \bibnamefont {Whaley}},\ }\bibfield  {title}
  {\bibinfo {title} {Geometric theory of nonlocal two-qubit operations},\
  }\href@noop {} {\bibfield  {journal} {\bibinfo  {journal} {Phys. Rev. A}\
  }\textbf {\bibinfo {volume} {67}},\ \bibinfo {pages} {042313} (\bibinfo
  {year} {2003})}\BibitemShut {NoStop}%
\bibitem [{\citenamefont {Ohfuchi}\ and\ \citenamefont
  {Sato}(2024)}]{remoteCR}%
  \BibitemOpen
  \bibfield  {author} {\bibinfo {author} {\bibfnamefont {M.}~\bibnamefont
  {Ohfuchi}}\ and\ \bibinfo {author} {\bibfnamefont {S.}~\bibnamefont {Sato}},\
  }\bibfield  {title} {\bibinfo {title} {Remote cross-resonance gate between
  superconducting fixed-frequency qubits},\ }\href@noop {} {\bibfield
  {journal} {\bibinfo  {journal} {Quantum Sci. Technol.}\ }\textbf {\bibinfo
  {volume} {9}},\ \bibinfo {pages} {035014} (\bibinfo {year}
  {2024})}\BibitemShut {NoStop}%
\bibitem [{\citenamefont {Jurcevic}\ \emph {et~al.}(2021)\citenamefont
  {Jurcevic}, \citenamefont {Javadi-Abhari}, \citenamefont {Bishop},
  \citenamefont {Lauer}, \citenamefont {Bogorin}, \citenamefont {Brink},
  \citenamefont {Capelluto}, \citenamefont {G{\"u}nl{\"u}k}, \citenamefont
  {Itoko}, \citenamefont {Kanazawa} \emph {et~al.}}]{QV64_IBM}%
  \BibitemOpen
  \bibfield  {author} {\bibinfo {author} {\bibfnamefont {P.}~\bibnamefont
  {Jurcevic}}, \bibinfo {author} {\bibfnamefont {A.}~\bibnamefont
  {Javadi-Abhari}}, \bibinfo {author} {\bibfnamefont {L.~S.}\ \bibnamefont
  {Bishop}}, \bibinfo {author} {\bibfnamefont {I.}~\bibnamefont {Lauer}},
  \bibinfo {author} {\bibfnamefont {D.~F.}\ \bibnamefont {Bogorin}}, \bibinfo
  {author} {\bibfnamefont {M.}~\bibnamefont {Brink}}, \bibinfo {author}
  {\bibfnamefont {L.}~\bibnamefont {Capelluto}}, \bibinfo {author}
  {\bibfnamefont {O.}~\bibnamefont {G{\"u}nl{\"u}k}}, \bibinfo {author}
  {\bibfnamefont {T.}~\bibnamefont {Itoko}}, \bibinfo {author} {\bibfnamefont
  {N.}~\bibnamefont {Kanazawa}}, \emph {et~al.},\ }\bibfield  {title} {\bibinfo
  {title} {Demonstration of quantum volume 64 on a superconducting quantum
  computing system},\ }\href@noop {} {\bibfield  {journal} {\bibinfo  {journal}
  {Quantum Science and Technology}\ }\textbf {\bibinfo {volume} {6}},\ \bibinfo
  {pages} {025020} (\bibinfo {year} {2021})}\BibitemShut {NoStop}%
\bibitem [{\citenamefont {de~Groot}\ \emph {et~al.}(2010)\citenamefont
  {de~Groot}, \citenamefont {Lisenfeld}, \citenamefont {Schouten},
  \citenamefont {Ashhab}, \citenamefont {Lupa{\c{s}}cu}, \citenamefont
  {Harmans},\ and\ \citenamefont {Mooij}}]{degroot_selectivedarkening}%
  \BibitemOpen
  \bibfield  {author} {\bibinfo {author} {\bibfnamefont {P.}~\bibnamefont
  {de~Groot}}, \bibinfo {author} {\bibfnamefont {J.}~\bibnamefont {Lisenfeld}},
  \bibinfo {author} {\bibfnamefont {R.}~\bibnamefont {Schouten}}, \bibinfo
  {author} {\bibfnamefont {S.}~\bibnamefont {Ashhab}}, \bibinfo {author}
  {\bibfnamefont {A.}~\bibnamefont {Lupa{\c{s}}cu}}, \bibinfo {author}
  {\bibfnamefont {C.}~\bibnamefont {Harmans}},\ and\ \bibinfo {author}
  {\bibfnamefont {J.}~\bibnamefont {Mooij}},\ }\bibfield  {title} {\bibinfo
  {title} {Selective darkening of degenerate transitions demonstrated with two
  superconducting quantum bits},\ }\href@noop {} {\bibfield  {journal}
  {\bibinfo  {journal} {Nat. Phys.}\ }\textbf {\bibinfo {volume} {6}},\
  \bibinfo {pages} {763} (\bibinfo {year} {2010})}\BibitemShut {NoStop}%
\bibitem [{\citenamefont {Rigetti}\ and\ \citenamefont
  {Devoret}(2010)}]{fullMW_regetti}%
  \BibitemOpen
  \bibfield  {author} {\bibinfo {author} {\bibfnamefont {C.}~\bibnamefont
  {Rigetti}}\ and\ \bibinfo {author} {\bibfnamefont {M.}~\bibnamefont
  {Devoret}},\ }\bibfield  {title} {\bibinfo {title} {Fully microwave-tunable
  universal gates in superconducting qubits with linear couplings and fixed
  transition frequencies},\ }\href@noop {} {\bibfield  {journal} {\bibinfo
  {journal} {Phys. Rev. B}\ }\textbf {\bibinfo {volume} {81}},\ \bibinfo
  {pages} {134507} (\bibinfo {year} {2010})}\BibitemShut {NoStop}%
\bibitem [{\citenamefont {Chow}\ \emph {et~al.}(2011)\citenamefont {Chow},
  \citenamefont {C{\'o}rcoles}, \citenamefont {Gambetta}, \citenamefont
  {Rigetti}, \citenamefont {Johnson}, \citenamefont {Smolin}, \citenamefont
  {Rozen}, \citenamefont {Keefe}, \citenamefont {Rothwell}, \citenamefont
  {Ketchen} \emph {et~al.}}]{chow_simpleAllMW}%
  \BibitemOpen
  \bibfield  {author} {\bibinfo {author} {\bibfnamefont {J.~M.}\ \bibnamefont
  {Chow}}, \bibinfo {author} {\bibfnamefont {A.~D.}\ \bibnamefont
  {C{\'o}rcoles}}, \bibinfo {author} {\bibfnamefont {J.~M.}\ \bibnamefont
  {Gambetta}}, \bibinfo {author} {\bibfnamefont {C.}~\bibnamefont {Rigetti}},
  \bibinfo {author} {\bibfnamefont {B.~R.}\ \bibnamefont {Johnson}}, \bibinfo
  {author} {\bibfnamefont {J.~A.}\ \bibnamefont {Smolin}}, \bibinfo {author}
  {\bibfnamefont {J.~R.}\ \bibnamefont {Rozen}}, \bibinfo {author}
  {\bibfnamefont {G.~A.}\ \bibnamefont {Keefe}}, \bibinfo {author}
  {\bibfnamefont {M.~B.}\ \bibnamefont {Rothwell}}, \bibinfo {author}
  {\bibfnamefont {M.~B.}\ \bibnamefont {Ketchen}}, \emph {et~al.},\ }\bibfield
  {title} {\bibinfo {title} {Simple all-microwave entangling gate for
  fixed-frequency superconducting qubits},\ }\href@noop {} {\bibfield
  {journal} {\bibinfo  {journal} {Phys. Rev. Lett.}\ }\textbf {\bibinfo
  {volume} {107}},\ \bibinfo {pages} {080502} (\bibinfo {year}
  {2011})}\BibitemShut {NoStop}%
\bibitem [{\citenamefont {de~Groot}\ \emph {et~al.}(2012)\citenamefont
  {de~Groot}, \citenamefont {Ashhab}, \citenamefont {Lupa{\c{s}}cu},
  \citenamefont {DiCarlo}, \citenamefont {Nori}, \citenamefont {Harmans},\ and\
  \citenamefont {Mooij}}]{degroot_selectivedarkening_2}%
  \BibitemOpen
  \bibfield  {author} {\bibinfo {author} {\bibfnamefont {P.}~\bibnamefont
  {de~Groot}}, \bibinfo {author} {\bibfnamefont {S.}~\bibnamefont {Ashhab}},
  \bibinfo {author} {\bibfnamefont {A.}~\bibnamefont {Lupa{\c{s}}cu}}, \bibinfo
  {author} {\bibfnamefont {L.}~\bibnamefont {DiCarlo}}, \bibinfo {author}
  {\bibfnamefont {F.}~\bibnamefont {Nori}}, \bibinfo {author} {\bibfnamefont
  {C.}~\bibnamefont {Harmans}},\ and\ \bibinfo {author} {\bibfnamefont
  {J.}~\bibnamefont {Mooij}},\ }\bibfield  {title} {\bibinfo {title} {Selective
  darkening of degenerate transitions for implementing quantum controlled-not
  gates},\ }\href@noop {} {\bibfield  {journal} {\bibinfo  {journal} {New J.
  Phys.}\ }\textbf {\bibinfo {volume} {14}},\ \bibinfo {pages} {073038}
  (\bibinfo {year} {2012})}\BibitemShut {NoStop}%
\bibitem [{\citenamefont {C{\'o}rcoles}\ \emph {et~al.}(2013)\citenamefont
  {C{\'o}rcoles}, \citenamefont {Gambetta}, \citenamefont {Chow}, \citenamefont
  {Smolin}, \citenamefont {Ware}, \citenamefont {Strand}, \citenamefont
  {Plourde},\ and\ \citenamefont {Steffen}}]{Corcoles_ProcessVerification}%
  \BibitemOpen
  \bibfield  {author} {\bibinfo {author} {\bibfnamefont {A.~D.}\ \bibnamefont
  {C{\'o}rcoles}}, \bibinfo {author} {\bibfnamefont {J.~M.}\ \bibnamefont
  {Gambetta}}, \bibinfo {author} {\bibfnamefont {J.~M.}\ \bibnamefont {Chow}},
  \bibinfo {author} {\bibfnamefont {J.~A.}\ \bibnamefont {Smolin}}, \bibinfo
  {author} {\bibfnamefont {M.}~\bibnamefont {Ware}}, \bibinfo {author}
  {\bibfnamefont {J.}~\bibnamefont {Strand}}, \bibinfo {author} {\bibfnamefont
  {B.~L.}\ \bibnamefont {Plourde}},\ and\ \bibinfo {author} {\bibfnamefont
  {M.}~\bibnamefont {Steffen}},\ }\bibfield  {title} {\bibinfo {title} {Process
  verification of two-qubit quantum gates by randomized benchmarking},\
  }\href@noop {} {\bibfield  {journal} {\bibinfo  {journal} {Phys. Rev. A}\
  }\textbf {\bibinfo {volume} {87}},\ \bibinfo {pages} {030301} (\bibinfo
  {year} {2013})}\BibitemShut {NoStop}%
\bibitem [{\citenamefont {Patterson}\ \emph {et~al.}(2019)\citenamefont
  {Patterson}, \citenamefont {Rahamim}, \citenamefont {Tsunoda}, \citenamefont
  {Spring}, \citenamefont {Jebari}, \citenamefont {Ratter}, \citenamefont
  {Mergenthaler}, \citenamefont {Tancredi}, \citenamefont {Vlastakis},
  \citenamefont {Esposito} \emph {et~al.}}]{Patterson_CaliCRgate}%
  \BibitemOpen
  \bibfield  {author} {\bibinfo {author} {\bibfnamefont {A.}~\bibnamefont
  {Patterson}}, \bibinfo {author} {\bibfnamefont {J.}~\bibnamefont {Rahamim}},
  \bibinfo {author} {\bibfnamefont {T.}~\bibnamefont {Tsunoda}}, \bibinfo
  {author} {\bibfnamefont {P.}~\bibnamefont {Spring}}, \bibinfo {author}
  {\bibfnamefont {S.}~\bibnamefont {Jebari}}, \bibinfo {author} {\bibfnamefont
  {K.}~\bibnamefont {Ratter}}, \bibinfo {author} {\bibfnamefont
  {M.}~\bibnamefont {Mergenthaler}}, \bibinfo {author} {\bibfnamefont
  {G.}~\bibnamefont {Tancredi}}, \bibinfo {author} {\bibfnamefont
  {B.}~\bibnamefont {Vlastakis}}, \bibinfo {author} {\bibfnamefont
  {M.}~\bibnamefont {Esposito}}, \emph {et~al.},\ }\bibfield  {title} {\bibinfo
  {title} {Calibration of a cross-resonance two-qubit gate between directly
  coupled transmons},\ }\href@noop {} {\bibfield  {journal} {\bibinfo
  {journal} {Phys. Rev. Appl.}\ }\textbf {\bibinfo {volume} {12}},\ \bibinfo
  {pages} {064013} (\bibinfo {year} {2019})}\BibitemShut {NoStop}%
\bibitem [{\citenamefont {Tripathi}\ \emph {et~al.}(2019)\citenamefont
  {Tripathi}, \citenamefont {Khezri},\ and\ \citenamefont
  {Korotkov}}]{Tripathi_OperationIntrinsicerrorbudget}%
  \BibitemOpen
  \bibfield  {author} {\bibinfo {author} {\bibfnamefont {V.}~\bibnamefont
  {Tripathi}}, \bibinfo {author} {\bibfnamefont {M.}~\bibnamefont {Khezri}},\
  and\ \bibinfo {author} {\bibfnamefont {A.~N.}\ \bibnamefont {Korotkov}},\
  }\bibfield  {title} {\bibinfo {title} {Operation and intrinsic error budget
  of a two-qubit cross-resonance gate},\ }\href@noop {} {\bibfield  {journal}
  {\bibinfo  {journal} {Phys. Rev. A}\ }\textbf {\bibinfo {volume} {100}},\
  \bibinfo {pages} {012301} (\bibinfo {year} {2019})}\BibitemShut {NoStop}%
\bibitem [{\citenamefont {Ware}\ \emph {et~al.}()\citenamefont {Ware},
  \citenamefont {Johnson}, \citenamefont {Gambetta}, \citenamefont {Ohki},
  \citenamefont {Chow},\ and\ \citenamefont {Plourde}}]{ware_cross}%
  \BibitemOpen
  \bibfield  {author} {\bibinfo {author} {\bibfnamefont {M.}~\bibnamefont
  {Ware}}, \bibinfo {author} {\bibfnamefont {B.}~\bibnamefont {Johnson}},
  \bibinfo {author} {\bibfnamefont {J.}~\bibnamefont {Gambetta}}, \bibinfo
  {author} {\bibfnamefont {T.}~\bibnamefont {Ohki}}, \bibinfo {author}
  {\bibfnamefont {J.}~\bibnamefont {Chow}},\ and\ \bibinfo {author}
  {\bibfnamefont {B.}~\bibnamefont {Plourde}},\ }\bibfield  {title} {\bibinfo
  {title} {Cross-resonance interactions between superconducting qubits with
  variable detuning},\ }\href@noop {} {\bibinfo  {journal} {Preprint at
  https://arxiv.org/abs/1905.11480}\ }\BibitemShut {NoStop}%
\bibitem [{\citenamefont {Ku}\ \emph {et~al.}(2020)\citenamefont {Ku},
  \citenamefont {Xu}, \citenamefont {Brink}, \citenamefont {McKay},
  \citenamefont {Hertzberg}, \citenamefont {Ansari},\ and\ \citenamefont
  {Plourde}}]{Ku_supperssionZZ}%
  \BibitemOpen
\bibfield  {journal} {  }\bibfield  {author} {\bibinfo {author} {\bibfnamefont
  {J.}~\bibnamefont {Ku}}, \bibinfo {author} {\bibfnamefont {X.}~\bibnamefont
  {Xu}}, \bibinfo {author} {\bibfnamefont {M.}~\bibnamefont {Brink}}, \bibinfo
  {author} {\bibfnamefont {D.~C.}\ \bibnamefont {McKay}}, \bibinfo {author}
  {\bibfnamefont {J.~B.}\ \bibnamefont {Hertzberg}}, \bibinfo {author}
  {\bibfnamefont {M.~H.}\ \bibnamefont {Ansari}},\ and\ \bibinfo {author}
  {\bibfnamefont {B.}~\bibnamefont {Plourde}},\ }\bibfield  {title} {\bibinfo
  {title} {Suppression of unwanted zz interactions in a hybrid two-qubit
  system},\ }\href@noop {} {\bibfield  {journal} {\bibinfo  {journal} {Phys.
  Rev. Lett.}\ }\textbf {\bibinfo {volume} {125}},\ \bibinfo {pages} {200504}
  (\bibinfo {year} {2020})}\BibitemShut {NoStop}%
\bibitem [{\citenamefont {Kandala}\ \emph {et~al.}(2021)\citenamefont
  {Kandala}, \citenamefont {Wei}, \citenamefont {Srinivasan}, \citenamefont
  {Magesan}, \citenamefont {Carnevale}, \citenamefont {Keefe}, \citenamefont
  {Klaus}, \citenamefont {Dial},\ and\ \citenamefont {McKay}}]{high_fidelity}%
  \BibitemOpen
  \bibfield  {author} {\bibinfo {author} {\bibfnamefont {A.}~\bibnamefont
  {Kandala}}, \bibinfo {author} {\bibfnamefont {K.~X.}\ \bibnamefont {Wei}},
  \bibinfo {author} {\bibfnamefont {S.}~\bibnamefont {Srinivasan}}, \bibinfo
  {author} {\bibfnamefont {E.}~\bibnamefont {Magesan}}, \bibinfo {author}
  {\bibfnamefont {S.}~\bibnamefont {Carnevale}}, \bibinfo {author}
  {\bibfnamefont {G.}~\bibnamefont {Keefe}}, \bibinfo {author} {\bibfnamefont
  {D.}~\bibnamefont {Klaus}}, \bibinfo {author} {\bibfnamefont
  {O.}~\bibnamefont {Dial}},\ and\ \bibinfo {author} {\bibfnamefont
  {D.}~\bibnamefont {McKay}},\ }\bibfield  {title} {\bibinfo {title}
  {Demonstration of a high-fidelity cnot gate for fixed-frequency transmons
  with engineered zz suppression},\ }\href@noop {} {\bibfield  {journal}
  {\bibinfo  {journal} {Phys. Rev. Lett.}\ }\textbf {\bibinfo {volume} {127}},\
  \bibinfo {pages} {130501} (\bibinfo {year} {2021})}\BibitemShut {NoStop}%
\bibitem [{\citenamefont {Malekakhlagh}\ \emph {et~al.}(2020)\citenamefont
  {Malekakhlagh}, \citenamefont {Magesan},\ and\ \citenamefont
  {McKay}}]{Malekakhlagh_firstprinciple_pra}%
  \BibitemOpen
  \bibfield  {author} {\bibinfo {author} {\bibfnamefont {M.}~\bibnamefont
  {Malekakhlagh}}, \bibinfo {author} {\bibfnamefont {E.}~\bibnamefont
  {Magesan}},\ and\ \bibinfo {author} {\bibfnamefont {D.~C.}\ \bibnamefont
  {McKay}},\ }\bibfield  {title} {\bibinfo {title} {First-principles analysis
  of cross-resonance gate operation},\ }\href@noop {} {\bibfield  {journal}
  {\bibinfo  {journal} {Phys. Rev. A}\ }\textbf {\bibinfo {volume} {102}},\
  \bibinfo {pages} {042605} (\bibinfo {year} {2020})}\BibitemShut {NoStop}%
\bibitem [{\citenamefont {Magesan}\ and\ \citenamefont
  {Gambetta}(2020)}]{megasan_effective_pra}%
  \BibitemOpen
  \bibfield  {author} {\bibinfo {author} {\bibfnamefont {E.}~\bibnamefont
  {Magesan}}\ and\ \bibinfo {author} {\bibfnamefont {J.~M.}\ \bibnamefont
  {Gambetta}},\ }\bibfield  {title} {\bibinfo {title} {Effective hamiltonian
  models of the cross-resonance gate},\ }\href@noop {} {\bibfield  {journal}
  {\bibinfo  {journal} {Phys. Rev. A}\ }\textbf {\bibinfo {volume} {101}},\
  \bibinfo {pages} {052308} (\bibinfo {year} {2020})}\BibitemShut {NoStop}%
\bibitem [{\citenamefont {Mitchell}\ \emph
  {et~al.}(2021{\natexlab{a}})\citenamefont {Mitchell}, \citenamefont {Naik},
  \citenamefont {Morvan}, \citenamefont {Hashim}, \citenamefont {Kreikebaum},
  \citenamefont {Marinelli}, \citenamefont {Lavrijsen}, \citenamefont
  {Nowrouzi}, \citenamefont {Santiago},\ and\ \citenamefont
  {Siddiqi}}]{mitchell_HardwareEfficient}%
  \BibitemOpen
  \bibfield  {author} {\bibinfo {author} {\bibfnamefont {B.~K.}\ \bibnamefont
  {Mitchell}}, \bibinfo {author} {\bibfnamefont {R.~K.}\ \bibnamefont {Naik}},
  \bibinfo {author} {\bibfnamefont {A.}~\bibnamefont {Morvan}}, \bibinfo
  {author} {\bibfnamefont {A.}~\bibnamefont {Hashim}}, \bibinfo {author}
  {\bibfnamefont {J.~M.}\ \bibnamefont {Kreikebaum}}, \bibinfo {author}
  {\bibfnamefont {B.}~\bibnamefont {Marinelli}}, \bibinfo {author}
  {\bibfnamefont {W.}~\bibnamefont {Lavrijsen}}, \bibinfo {author}
  {\bibfnamefont {K.}~\bibnamefont {Nowrouzi}}, \bibinfo {author}
  {\bibfnamefont {D.~I.}\ \bibnamefont {Santiago}},\ and\ \bibinfo {author}
  {\bibfnamefont {I.}~\bibnamefont {Siddiqi}},\ }\bibfield  {title} {\bibinfo
  {title} {Hardware-efficient microwave-activated tunable coupling between
  superconducting qubits},\ }\href@noop {} {\bibfield  {journal} {\bibinfo
  {journal} {Phys. Rev. Lett.}\ }\textbf {\bibinfo {volume} {127}},\ \bibinfo
  {pages} {200502} (\bibinfo {year} {2021}{\natexlab{a}})}\BibitemShut
  {NoStop}%
\bibitem [{\citenamefont {Wei}\ \emph {et~al.}(2021)\citenamefont {Wei},
  \citenamefont {Magesan}, \citenamefont {Lauer}, \citenamefont {Srinivasan},
  \citenamefont {Bogorin}, \citenamefont {Carnevale}, \citenamefont {Keefe},
  \citenamefont {Kim}, \citenamefont {Klaus}, \citenamefont {Landers} \emph
  {et~al.}}]{Wei_CrosstalkCancellation}%
  \BibitemOpen
  \bibfield  {author} {\bibinfo {author} {\bibfnamefont {K.}~\bibnamefont
  {Wei}}, \bibinfo {author} {\bibfnamefont {E.}~\bibnamefont {Magesan}},
  \bibinfo {author} {\bibfnamefont {I.}~\bibnamefont {Lauer}}, \bibinfo
  {author} {\bibfnamefont {S.}~\bibnamefont {Srinivasan}}, \bibinfo {author}
  {\bibfnamefont {D.}~\bibnamefont {Bogorin}}, \bibinfo {author} {\bibfnamefont
  {S.}~\bibnamefont {Carnevale}}, \bibinfo {author} {\bibfnamefont
  {G.}~\bibnamefont {Keefe}}, \bibinfo {author} {\bibfnamefont
  {Y.}~\bibnamefont {Kim}}, \bibinfo {author} {\bibfnamefont {D.}~\bibnamefont
  {Klaus}}, \bibinfo {author} {\bibfnamefont {W.}~\bibnamefont {Landers}},
  \emph {et~al.},\ }\bibfield  {title} {\bibinfo {title} {Quantum crosstalk
  cancellation for fast entangling gates and improved multi-qubit
  performance},\ }\href@noop {} {\bibfield  {journal} {\bibinfo  {journal}
  {Preprint at https://arxiv.org/abs/2106.00675}\ } (\bibinfo {year}
  {2021})}\BibitemShut {NoStop}%
\bibitem [{\citenamefont {Xu}\ and\ \citenamefont
  {Ansari}(2021)}]{xu_zzfreedom}%
  \BibitemOpen
  \bibfield  {author} {\bibinfo {author} {\bibfnamefont {X.}~\bibnamefont
  {Xu}}\ and\ \bibinfo {author} {\bibfnamefont {M.}~\bibnamefont {Ansari}},\
  }\bibfield  {title} {\bibinfo {title} {Zz freedom in two-qubit gates},\
  }\href@noop {} {\bibfield  {journal} {\bibinfo  {journal} {Phys. Rev. Appl.}\
  }\textbf {\bibinfo {volume} {15}},\ \bibinfo {pages} {064074} (\bibinfo
  {year} {2021})}\BibitemShut {NoStop}%
\bibitem [{\citenamefont {Mitchell}\ \emph
  {et~al.}(2021{\natexlab{b}})\citenamefont {Mitchell}, \citenamefont {Naik},
  \citenamefont {Morvan}, \citenamefont {Hashim}, \citenamefont {Kreikebaum},
  \citenamefont {Marinelli}, \citenamefont {Lavrijsen}, \citenamefont
  {Nowrouzi}, \citenamefont {Santiago},\ and\ \citenamefont
  {Siddiqi}}]{mitchell2021hardware}%
  \BibitemOpen
  \bibfield  {author} {\bibinfo {author} {\bibfnamefont {B.~K.}\ \bibnamefont
  {Mitchell}}, \bibinfo {author} {\bibfnamefont {R.~K.}\ \bibnamefont {Naik}},
  \bibinfo {author} {\bibfnamefont {A.}~\bibnamefont {Morvan}}, \bibinfo
  {author} {\bibfnamefont {A.}~\bibnamefont {Hashim}}, \bibinfo {author}
  {\bibfnamefont {J.~M.}\ \bibnamefont {Kreikebaum}}, \bibinfo {author}
  {\bibfnamefont {B.}~\bibnamefont {Marinelli}}, \bibinfo {author}
  {\bibfnamefont {W.}~\bibnamefont {Lavrijsen}}, \bibinfo {author}
  {\bibfnamefont {K.}~\bibnamefont {Nowrouzi}}, \bibinfo {author}
  {\bibfnamefont {D.~I.}\ \bibnamefont {Santiago}},\ and\ \bibinfo {author}
  {\bibfnamefont {I.}~\bibnamefont {Siddiqi}},\ }\bibfield  {title} {\bibinfo
  {title} {Hardware-efficient microwave-activated tunable coupling between
  superconducting qubits},\ }\href
  {https://doi.org/10.1103/PhysRevLett.127.200502} {\bibfield  {journal}
  {\bibinfo  {journal} {Phys. Rev. Lett.}\ }\textbf {\bibinfo {volume} {127}},\
  \bibinfo {pages} {200502} (\bibinfo {year} {2021}{\natexlab{b}})}\BibitemShut
  {NoStop}%
\bibitem [{\citenamefont {Wei}\ \emph {et~al.}(2022)\citenamefont {Wei},
  \citenamefont {Magesan}, \citenamefont {Lauer}, \citenamefont {Srinivasan},
  \citenamefont {Bogorin}, \citenamefont {Carnevale}, \citenamefont {Keefe},
  \citenamefont {Kim}, \citenamefont {Klaus}, \citenamefont {Landers},
  \citenamefont {Sundaresan}, \citenamefont {Wang}, \citenamefont {Zhang},
  \citenamefont {Steffen}, \citenamefont {Dial}, \citenamefont {McKay},\ and\
  \citenamefont {Kandala}}]{wei2022hamiltonian}%
  \BibitemOpen
  \bibfield  {author} {\bibinfo {author} {\bibfnamefont {K.~X.}\ \bibnamefont
  {Wei}}, \bibinfo {author} {\bibfnamefont {E.}~\bibnamefont {Magesan}},
  \bibinfo {author} {\bibfnamefont {I.}~\bibnamefont {Lauer}}, \bibinfo
  {author} {\bibfnamefont {S.}~\bibnamefont {Srinivasan}}, \bibinfo {author}
  {\bibfnamefont {D.~F.}\ \bibnamefont {Bogorin}}, \bibinfo {author}
  {\bibfnamefont {S.}~\bibnamefont {Carnevale}}, \bibinfo {author}
  {\bibfnamefont {G.~A.}\ \bibnamefont {Keefe}}, \bibinfo {author}
  {\bibfnamefont {Y.}~\bibnamefont {Kim}}, \bibinfo {author} {\bibfnamefont
  {D.}~\bibnamefont {Klaus}}, \bibinfo {author} {\bibfnamefont
  {W.}~\bibnamefont {Landers}}, \bibinfo {author} {\bibfnamefont
  {N.}~\bibnamefont {Sundaresan}}, \bibinfo {author} {\bibfnamefont
  {C.}~\bibnamefont {Wang}}, \bibinfo {author} {\bibfnamefont {E.~J.}\
  \bibnamefont {Zhang}}, \bibinfo {author} {\bibfnamefont {M.}~\bibnamefont
  {Steffen}}, \bibinfo {author} {\bibfnamefont {O.~E.}\ \bibnamefont {Dial}},
  \bibinfo {author} {\bibfnamefont {D.~C.}\ \bibnamefont {McKay}},\ and\
  \bibinfo {author} {\bibfnamefont {A.}~\bibnamefont {Kandala}},\ }\bibfield
  {title} {\bibinfo {title} {Hamiltonian engineering with multicolor drives for
  fast entangling gates and quantum crosstalk cancellation},\ }\href
  {https://doi.org/10.1103/PhysRevLett.129.060501} {\bibfield  {journal}
  {\bibinfo  {journal} {Phys. Rev. Lett.}\ }\textbf {\bibinfo {volume} {129}},\
  \bibinfo {pages} {060501} (\bibinfo {year} {2022})}\BibitemShut {NoStop}%
\bibitem [{\citenamefont {Qiu}\ \emph {et~al.}(2023)\citenamefont {Qiu},
  \citenamefont {Liu}, \citenamefont {Niu}, \citenamefont {Hu}, \citenamefont
  {Wu}, \citenamefont {Zhang}, \citenamefont {Huang}, \citenamefont {Chen},
  \citenamefont {Li}, \citenamefont {Liu} \emph
  {et~al.}}]{deterministictele_zhong}%
  \BibitemOpen
  \bibfield  {author} {\bibinfo {author} {\bibfnamefont {J.}~\bibnamefont
  {Qiu}}, \bibinfo {author} {\bibfnamefont {Y.}~\bibnamefont {Liu}}, \bibinfo
  {author} {\bibfnamefont {J.}~\bibnamefont {Niu}}, \bibinfo {author}
  {\bibfnamefont {L.}~\bibnamefont {Hu}}, \bibinfo {author} {\bibfnamefont
  {Y.}~\bibnamefont {Wu}}, \bibinfo {author} {\bibfnamefont {L.}~\bibnamefont
  {Zhang}}, \bibinfo {author} {\bibfnamefont {W.}~\bibnamefont {Huang}},
  \bibinfo {author} {\bibfnamefont {Y.}~\bibnamefont {Chen}}, \bibinfo {author}
  {\bibfnamefont {J.}~\bibnamefont {Li}}, \bibinfo {author} {\bibfnamefont
  {S.}~\bibnamefont {Liu}}, \emph {et~al.},\ }\bibfield  {title} {\bibinfo
  {title} {Deterministic quantum teleportation between distant superconducting
  chips},\ }\href@noop {} {\bibfield  {journal} {\bibinfo  {journal} {Preprint
  at https://arxiv.org/abs/2302.08756}\ } (\bibinfo {year} {2023})}\BibitemShut
  {NoStop}%
\bibitem [{\citenamefont {Sheldon}\ \emph {et~al.}(2016)\citenamefont
  {Sheldon}, \citenamefont {Magesan}, \citenamefont {Chow},\ and\ \citenamefont
  {Gambetta}}]{CRparameters_sheldon}%
  \BibitemOpen
  \bibfield  {author} {\bibinfo {author} {\bibfnamefont {S.}~\bibnamefont
  {Sheldon}}, \bibinfo {author} {\bibfnamefont {E.}~\bibnamefont {Magesan}},
  \bibinfo {author} {\bibfnamefont {J.~M.}\ \bibnamefont {Chow}},\ and\
  \bibinfo {author} {\bibfnamefont {J.~M.}\ \bibnamefont {Gambetta}},\
  }\bibfield  {title} {\bibinfo {title} {Procedure for systematically tuning up
  cross-talk in the cross-resonance gate},\ }\href@noop {} {\bibfield
  {journal} {\bibinfo  {journal} {Phys. Rev. A}\ }\textbf {\bibinfo {volume}
  {93}},\ \bibinfo {pages} {060302} (\bibinfo {year} {2016})}\BibitemShut
  {NoStop}%
\bibitem [{\citenamefont {Neill}\ \emph {et~al.}(2018)\citenamefont {Neill},
  \citenamefont {Roushan}, \citenamefont {Kechedzhi}, \citenamefont {Boixo},
  \citenamefont {Isakov}, \citenamefont {Smelyanskiy}, \citenamefont {Megrant},
  \citenamefont {Chiaro}, \citenamefont {Dunsworth}, \citenamefont {Arya} \emph
  {et~al.}}]{blueprint_Martinis}%
  \BibitemOpen
  \bibfield  {author} {\bibinfo {author} {\bibfnamefont {C.}~\bibnamefont
  {Neill}}, \bibinfo {author} {\bibfnamefont {P.}~\bibnamefont {Roushan}},
  \bibinfo {author} {\bibfnamefont {K.}~\bibnamefont {Kechedzhi}}, \bibinfo
  {author} {\bibfnamefont {S.}~\bibnamefont {Boixo}}, \bibinfo {author}
  {\bibfnamefont {S.~V.}\ \bibnamefont {Isakov}}, \bibinfo {author}
  {\bibfnamefont {V.}~\bibnamefont {Smelyanskiy}}, \bibinfo {author}
  {\bibfnamefont {A.}~\bibnamefont {Megrant}}, \bibinfo {author} {\bibfnamefont
  {B.}~\bibnamefont {Chiaro}}, \bibinfo {author} {\bibfnamefont
  {A.}~\bibnamefont {Dunsworth}}, \bibinfo {author} {\bibfnamefont
  {K.}~\bibnamefont {Arya}}, \emph {et~al.},\ }\bibfield  {title} {\bibinfo
  {title} {A blueprint for demonstrating quantum supremacy with superconducting
  qubits},\ }\href@noop {} {\bibfield  {journal} {\bibinfo  {journal}
  {Science}\ }\textbf {\bibinfo {volume} {360}},\ \bibinfo {pages} {195}
  (\bibinfo {year} {2018})}\BibitemShut {NoStop}%
\bibitem [{\citenamefont {Clauser}\ \emph {et~al.}(1969)\citenamefont
  {Clauser}, \citenamefont {Horne}, \citenamefont {Shimony},\ and\
  \citenamefont {Holt}}]{CHSH}%
  \BibitemOpen
  \bibfield  {author} {\bibinfo {author} {\bibfnamefont {J.~F.}\ \bibnamefont
  {Clauser}}, \bibinfo {author} {\bibfnamefont {M.~A.}\ \bibnamefont {Horne}},
  \bibinfo {author} {\bibfnamefont {A.}~\bibnamefont {Shimony}},\ and\ \bibinfo
  {author} {\bibfnamefont {R.~A.}\ \bibnamefont {Holt}},\ }\bibfield  {title}
  {\bibinfo {title} {Proposed experiment to test local hidden-variable
  theories},\ }\href {https://doi.org/10.1103/PhysRevLett.23.880} {\bibfield
  {journal} {\bibinfo  {journal} {Phys. Rev. Lett.}\ }\textbf {\bibinfo
  {volume} {23}},\ \bibinfo {pages} {880} (\bibinfo {year} {1969})}\BibitemShut
  {NoStop}%
\bibitem [{\citenamefont {Mollenhauer}\ \emph {et~al.}(2024)\citenamefont
  {Mollenhauer}, \citenamefont {Irfan}, \citenamefont {Cao}, \citenamefont
  {Mandal},\ and\ \citenamefont {Pfaff}}]{mollenhauer2024high}%
  \BibitemOpen
  \bibfield  {author} {\bibinfo {author} {\bibfnamefont {M.}~\bibnamefont
  {Mollenhauer}}, \bibinfo {author} {\bibfnamefont {A.}~\bibnamefont {Irfan}},
  \bibinfo {author} {\bibfnamefont {X.}~\bibnamefont {Cao}}, \bibinfo {author}
  {\bibfnamefont {S.}~\bibnamefont {Mandal}},\ and\ \bibinfo {author}
  {\bibfnamefont {W.}~\bibnamefont {Pfaff}},\ }\bibfield  {title} {\bibinfo
  {title} {A high-efficiency plug-and-play superconducting qubit network},\
  }\href@noop {} {\bibfield  {journal} {\bibinfo  {journal} {Preprint at
  https://arxiv.org/abs/2407.16743}\ } (\bibinfo {year} {2024})}\BibitemShut
  {NoStop}%
\bibitem [{\citenamefont {Dolan}(1977)}]{dolan1977offset}%
  \BibitemOpen
  \bibfield  {author} {\bibinfo {author} {\bibfnamefont {G.~J.}\ \bibnamefont
  {Dolan}},\ }\bibfield  {title} {\bibinfo {title} {{Offset masks for
  lift‐off photoprocessing}},\ }\href {https://doi.org/10.1063/1.89690}
  {\bibfield  {journal} {\bibinfo  {journal} {Appl. Phys. Lett.}\ }\textbf
  {\bibinfo {volume} {31}},\ \bibinfo {pages} {337} (\bibinfo {year}
  {1977})}\BibitemShut {NoStop}%
\bibitem [{\citenamefont {Chen}\ \emph {et~al.}(2016)\citenamefont {Chen},
  \citenamefont {Kelly}, \citenamefont {Quintana}, \citenamefont {Barends},
  \citenamefont {Campbell}, \citenamefont {Chen}, \citenamefont {Chiaro},
  \citenamefont {Dunsworth}, \citenamefont {Fowler}, \citenamefont {Lucero}
  \emph {et~al.}}]{DRAG_exp}%
  \BibitemOpen
  \bibfield  {author} {\bibinfo {author} {\bibfnamefont {Z.}~\bibnamefont
  {Chen}}, \bibinfo {author} {\bibfnamefont {J.}~\bibnamefont {Kelly}},
  \bibinfo {author} {\bibfnamefont {C.}~\bibnamefont {Quintana}}, \bibinfo
  {author} {\bibfnamefont {R.}~\bibnamefont {Barends}}, \bibinfo {author}
  {\bibfnamefont {B.}~\bibnamefont {Campbell}}, \bibinfo {author}
  {\bibfnamefont {Y.}~\bibnamefont {Chen}}, \bibinfo {author} {\bibfnamefont
  {B.}~\bibnamefont {Chiaro}}, \bibinfo {author} {\bibfnamefont
  {A.}~\bibnamefont {Dunsworth}}, \bibinfo {author} {\bibfnamefont
  {A.}~\bibnamefont {Fowler}}, \bibinfo {author} {\bibfnamefont
  {E.}~\bibnamefont {Lucero}}, \emph {et~al.},\ }\bibfield  {title} {\bibinfo
  {title} {Measuring and suppressing quantum state leakage in a superconducting
  qubit},\ }\href@noop {} {\bibfield  {journal} {\bibinfo  {journal} {Phys.
  Rev. Lett.}\ }\textbf {\bibinfo {volume} {116}},\ \bibinfo {pages} {020501}
  (\bibinfo {year} {2016})}\BibitemShut {NoStop}%
\bibitem [{\citenamefont {Gambetta}\ \emph {et~al.}(2011)\citenamefont
  {Gambetta}, \citenamefont {Motzoi}, \citenamefont {Merkel},\ and\
  \citenamefont {Wilhelm}}]{DRAG_theo}%
  \BibitemOpen
  \bibfield  {author} {\bibinfo {author} {\bibfnamefont {J.~M.}\ \bibnamefont
  {Gambetta}}, \bibinfo {author} {\bibfnamefont {F.}~\bibnamefont {Motzoi}},
  \bibinfo {author} {\bibfnamefont {S.}~\bibnamefont {Merkel}},\ and\ \bibinfo
  {author} {\bibfnamefont {F.~K.}\ \bibnamefont {Wilhelm}},\ }\bibfield
  {title} {\bibinfo {title} {Analytic control methods for high-fidelity unitary
  operations in a weakly nonlinear oscillator},\ }\href@noop {} {\bibfield
  {journal} {\bibinfo  {journal} {Phys. Rev. A}\ }\textbf {\bibinfo {volume}
  {83}},\ \bibinfo {pages} {012308} (\bibinfo {year} {2011})}\BibitemShut
  {NoStop}%
\bibitem [{\citenamefont {Shen}\ and\ \citenamefont
  {Duan}(2012)}]{shen2012correcting}%
  \BibitemOpen
  \bibfield  {author} {\bibinfo {author} {\bibfnamefont {C.}~\bibnamefont
  {Shen}}\ and\ \bibinfo {author} {\bibfnamefont {L.}~\bibnamefont {Duan}},\
  }\bibfield  {title} {\bibinfo {title} {Correcting detection errors in quantum
  state engineering through data processing},\ }\href@noop {} {\bibfield
  {journal} {\bibinfo  {journal} {New Journal of Physics}\ }\textbf {\bibinfo
  {volume} {14}},\ \bibinfo {pages} {053053} (\bibinfo {year}
  {2012})}\BibitemShut {NoStop}%
\bibitem [{\citenamefont {Nation}\ \emph {et~al.}(2021)\citenamefont {Nation},
  \citenamefont {Kang}, \citenamefont {Sundaresan},\ and\ \citenamefont
  {Gambetta}}]{nation2021scalable}%
  \BibitemOpen
  \bibfield  {author} {\bibinfo {author} {\bibfnamefont {P.~D.}\ \bibnamefont
  {Nation}}, \bibinfo {author} {\bibfnamefont {H.}~\bibnamefont {Kang}},
  \bibinfo {author} {\bibfnamefont {N.}~\bibnamefont {Sundaresan}},\ and\
  \bibinfo {author} {\bibfnamefont {J.~M.}\ \bibnamefont {Gambetta}},\
  }\bibfield  {title} {\bibinfo {title} {Scalable mitigation of measurement
  errors on quantum computers},\ }\href@noop {} {\bibfield  {journal} {\bibinfo
   {journal} {PRX Quantum}\ }\textbf {\bibinfo {volume} {2}},\ \bibinfo {pages}
  {040326} (\bibinfo {year} {2021})}\BibitemShut {NoStop}%
\end{thebibliography}%

\section{Acknowledgements}
We thank Dr. He Wang and Dr. Kehuan Linghu for their helpful discussions. We acknowledge supports from the National Natural Science Foundation of China (Grant Nos. 92365206, 12322413, 12404557, 12404560), Innovation Program for Quantum Science and Technology (No.2021ZD0301802).

\section{Author contributions}

W. Z. and F. Y. conceived the experiment; J. S., G. X., and Z. M. designed and fabricated the quantum device; S. Y. and W. Z. performed the measurement; P. L. analyzed the experimental data and performed the simulation; P. L. and W. Z. wrote the manuscript; F. Y., Y. J., and H. Y. supervised on the experiment; All the authors discussed on the results and the manuscript.

\newpage

\section{Methods}

\subsection{Sample fabrication}

The circuit layer and Josephson junctions are fabricated on a single-side polished sapphire substrate. The circuit consists of a transmission line, a readout resonator, and a half-wavelength co-planar waveguide (HW-CPW). The fabrication process is as follows.

1. The substrate is annealed in the air at 1200 $^\circ$C for 3 hours.

2. 200-nm $\alpha$-tantalum ($\alpha$-Ta) film is deposited at 500 $^\circ$C by magnetron sputtering.

3. The substrate is spin-coated with S1813 photoresist. The base circuit is then realized through direct write lithography (by Heidelberg DWL 66$+$). After exposure, the pattern is first developed by MF 319 developer, and then ${\rm CF}_4$ reactive ion etching (RIE) is employed to remove exposed Ta layer.

4. The photoresist is subsequently removed by ultrasonic treatment in acetone, then the sample is immersed in a piranha solution (a 2:1 volume ratio mixture of ${\rm H}_2{\rm SO}_4$ and ${\rm H}_2{\rm O}_2$) for 20 minutes to eliminate residual photoresist.

5. Utilizing electron beam lithography (EBL) with a bi-layer of photoresists (PMMA A4/LOR 10B), the Josephson junction patterns are precisely exposed and subsequently developed by a mixture of MIBK and IPA solution (${\rm MIBK}:{\rm IPA}=1:3$) and then MF 319 solution, respectively. 

6. To ensure the superconducting connection between Ta and Al, the Ta oxides and residual resists on Josephson junction patterns are removed via ion beam etching before the deposition of Al electrode. The Josephson junctions are fabricated by Dolan bridge technique \cite{dolan1977offset}.

7. Once the wafer is diced into individual chips, the lift-off process is executed in N-Methylpyrrolidone. The sample is subsequently wirebonded to a copper sample box using the superconducting Al wire (See FIG. \ref{fig:supp_sample}).

\subsection{Connection between chip and cable}

\begin{figure}[htbp]
    \centering
    \includegraphics[]{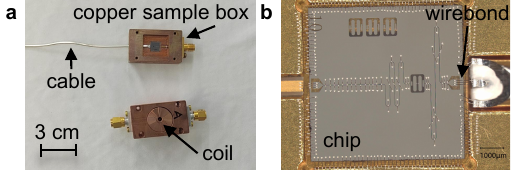}
    \caption{\textbf{a}, Image of the device consisting of a copper sample box, a chip, and an Al coaxial cable. \textbf{b}, Photograph of a quantum module.}
    \label{fig:supp_sample}
\end{figure}

\begin{figure}[htbp]
    \centering
    \includegraphics[]{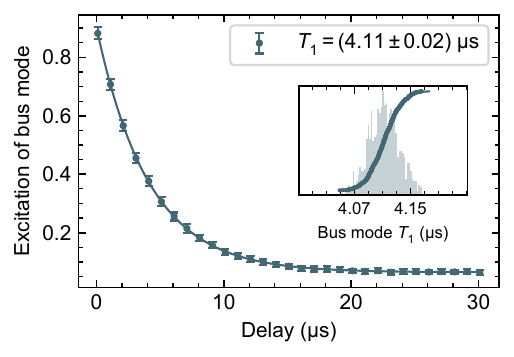}
    \caption{\textbf{Life time of the bus coupling mode}. The inset is the histogram of bus mode $T_1$ calculated by bootstrapping.}
    \label{fig:bus_life_time}
\end{figure}

In this experiment, two $7\times7\text{ mm}^2$ chips are placed within two copper sample boxes. The two modules are connected by the 30-cm-long Al coaxial cable (provided by Nanjing Hermerc System). The cable is fixed to the sample box by solder. The HW-CPW is connected to the Al coaxial cable via superconducting Al wirebonds. The outer conductors and the dielectric material at both ends of the cable are removed to expose the inner conductors for wire bonding, as displayed in FIG. \ref{fig:supp_sample}b.

According to the information provided by the manufacturer of Al coaxial line, the outer diameter of the inner conductor of the coaxial line is $0.54$ mm and the inner diameter of the outer conductor is $2.05$ mm. The insulator is Polytetrafluoroethylene (PTFE) whose dielectric constant is about $2.55$. The frequency distance between each longitudinal modes are designed to be $273.31$ MHz.

To measure the lifetime of the transmission line, we utilize the $|0,f\rangle \leftrightarrow |1,g\rangle$ transition to inject a MW photon into the bus mode. Then, we extract the photon from the transmission line after a period of delay. By fitting the residual photon inside the transmission line vs the delay period, we get the photon lifetime $T_1 = (4.11 \pm 0.02) \mu s$. The standing wave mode we used for coupling is $\omega/ 2\pi = 4.586$ GHz. Thus, the Q-factor of this mode is calculated to be $Q=\omega T_1=1.18 \times 10 ^{5}$ (FIG. \ref{fig:bus_life_time}).

\subsection{Experimental setup}

\begin{figure}[htbp]
    \centering
    \includegraphics[width=8.64cm]{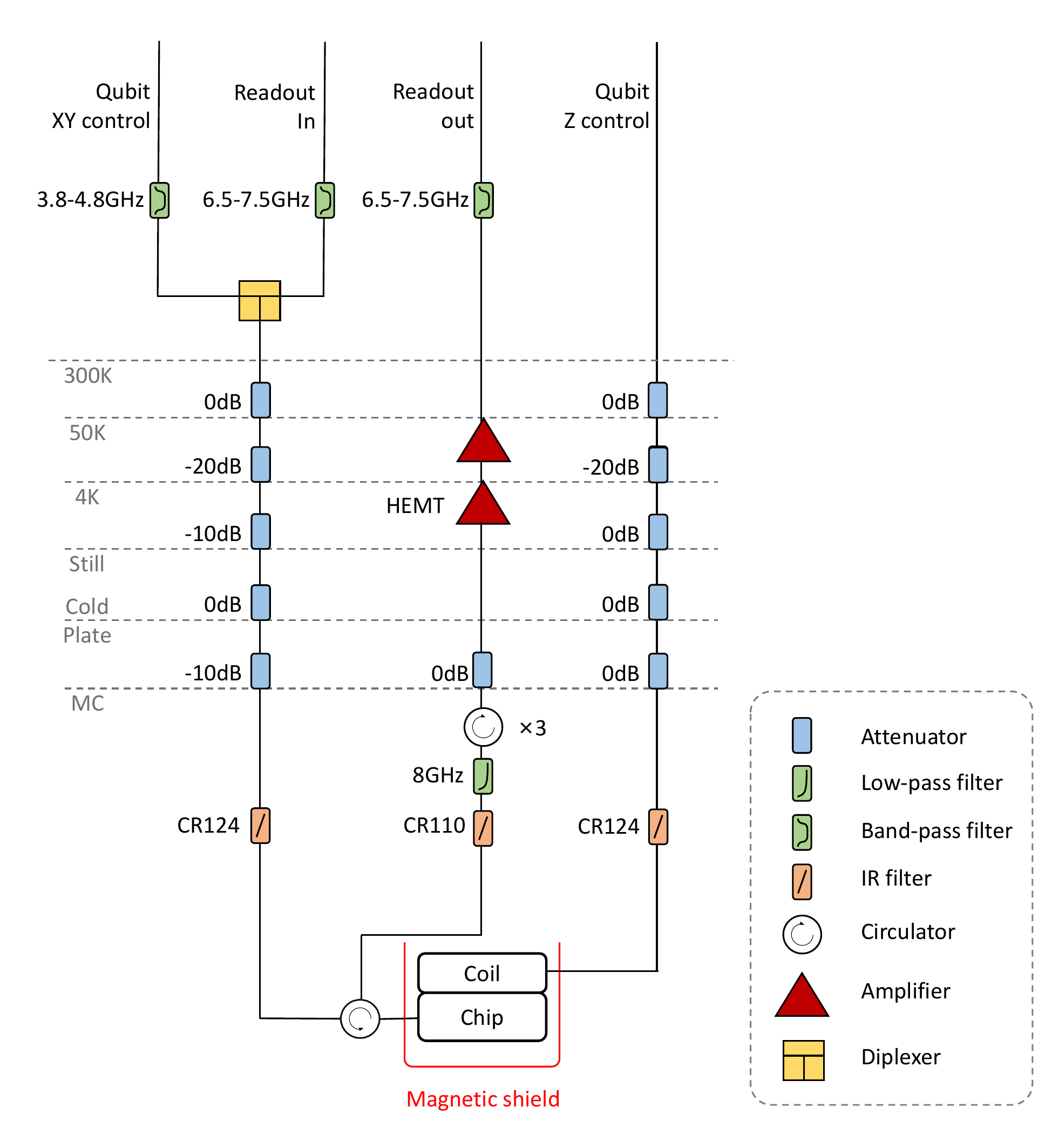}
    \caption{\textbf{Control and measurement system for a package}. The system includes the cryogenic and room-temperature setup.}
    \label{fig:fridge}
\end{figure}

It shows the electronic system we used in our experiment in FIG. \ref{fig:fridge}. For each qubit, the XY-control signal and readout signal are combined with a diplexer at room temperature. The XY control and readout input signals are generated by a direct digital synthesis (DDS, 6 GHz sampling rate). The readout output signals are captured directly by the analog-to-digital converters (ADC, 4GHz sampling rate) on the same board. Two high-electron-mobility transistors (HEMT) are used for output signal amplification (Connphy TCA 2690 at 4K, HMC system AMPG00811A001 at 50K). The Z control signals are generated by an arbitrary wave generator (AWG, 2 GHz sampling rate).

\subsection{Qubit and Coupling Information}

\begin{figure*}[htbp]
    \centering
    \includegraphics[]{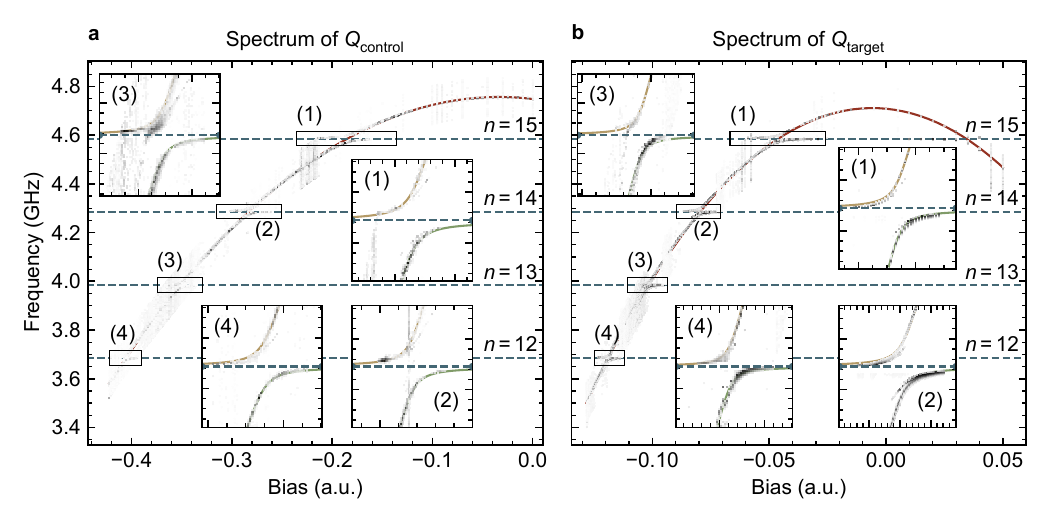}
    \caption{\textbf{Spectrum of each qubit}. As the bias varying, i.e. the frequency of each qubit changing, the interaction between each qubit and four bus modes can be found. The frequencies of four modes $n=15$, $n=14$, $n=13$, and $n=12$ are labeled by the blue dashed lines. The insets illustrate details of the anti-cross region in the spectrum, correspondingly. The experimental spectrum data is in gray and colorful lines are the fitting results.}
    \label{fig:spectrum}
\end{figure*}

\begin{table}[htbp]
    \centering
    \begin{tabular}{c|ccc}
    \toprule
    Mode&$\omega_n/2\pi$ (GHz)&$g_{c,n}/2 \pi$ (MHz)&$g_{t,n}/2\pi$ (MHz)\\
    \hline
    $n=15$&4.586&14.9&14.0\\
    $n=14$&4.286&13.2&12.5\\
    $n=13$&3.984&12.2&11.7\\
    $n=12$&3.686&11.6&10.4\\
    \hline
    $n=16$&4.886&15.7&15.2\\
    \hline
    \end{tabular}
    \caption{\textbf{Coupling information}. Frequency of each bus mode $\omega_n/2\pi$ extracted from spectrum data are listed, as well as the coupling strength between the bus mode and the control (target) qubit $g_{c,n}/2\pi$ ($g_{t,n}/2\pi$). The frequency and coupling strengths of $n=16$ are inferred from the other four mode by a linear fitting, which is used in the simulations.}
    \label{tab:coupling_information}
\end{table}

\begin{table}[htbp]
    \centering
    \begin{tabular}{c|cc}
    \toprule
         & $Q_c$ & $Q_t$\\
    \hline
        transition frequency $\omega_{01}/2\pi$ (GHz) & 4.743 & 4.698 \\
        anharmonicity $\alpha /2\pi$ (MHz)& -237 & -226 \\
        readout frequency $\omega_{r}/2\pi$ (GHz)& 6.920 & 6.920  \\
        lifetime $T_1$ ($\mu$s)& 41.6 & 59.0  \\
        dephasing time $T_2^{\rm Ramsey}$ ($\mu$s)& 7.93 & 8.30\\
        dephasing time $T_2^{\rm Echo}$ ($\mu$s) & 9.08 & 8.77 \\
        dephasing time $T_2^{\rm XY8}$ ($\mu$s)  & 21.4 & 22.4 \\
    \hline
    \end{tabular}
    \caption{\textbf{Qubit information}. Frequency and coherence information of qubits at the working point. The qubit-bus coupling strength is measured using the energy level anti-cross at the 15th standing mode of the coupling bus.}
    \label{tab:qubits_information}
\end{table}

\begin{figure}[htbp]
    \centering
    \includegraphics[]{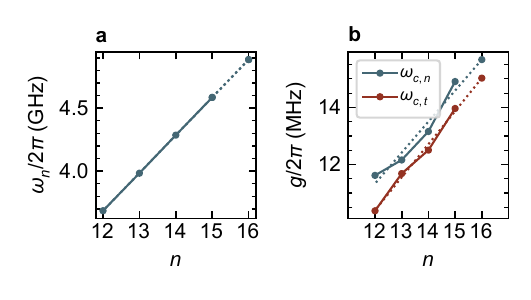}
    \caption{\textbf{Bus mode frequency and coupling}. The data is listed in TABLE \ref{tab:coupling_information}. The dotted lines are the linear extrapolations.}
    \label{fig:anti_cross}
\end{figure}

The spectra of both qubits as the magnetic flux bias varying are shown in FIG. \ref{fig:spectrum}. The anti-crossing effect can be found when the qubit frequency is close to the bus mode frequency. Here, the anti-crossing between the (1) $n=15$, (2) $n=14$, (3) $n=13$, and (4) $n=12$ are observed and shown in the insets of FIG. \ref{fig:spectrum}, respectively. According to the spectrum fitting, the coupling strength between the $n$th bus mode and the qubit can be extracted, which is listed and shown in TABLE \ref{tab:coupling_information} and FIG. \ref{fig:anti_cross}. The bus mode frequency $\omega_n$ appears in a linearity, thus, we can infer the frequency of $n=16$ mode via a linear fitting extrapolation. As for the coupling strength $g_{c, n}$ ($g_{t, n}$), it is roughly estimated to be proportional to $\sqrt{\omega_n\omega_c}$ ($\sqrt{\omega_n\omega_t}$). Therefore, the linear extrapolation for $g_{c, n}$ ($g_{t, n}$) is also valid when $n=16$. However, the frequency splitting on the anti-crossing region is sensitive to other coupling channels so that we have to pay attention to the measurement deviation.

In TABLE \ref{tab:qubits_information} we display the characteristics of each qubit at working bias in gate calibrations. Besides the reason why we select this working bias mentioned in the main text, coherence time of qubits is another point to determine the working bias.

\subsection{Effective CR Hamiltonian and Simulations}

\begin{figure}[htbp]
    \centering
    \includegraphics[]{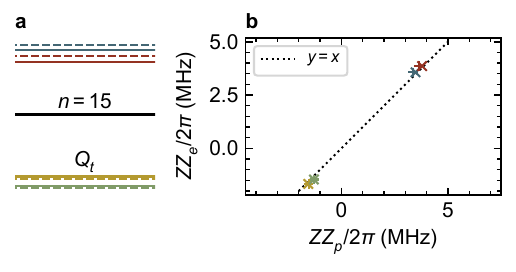}
    \caption{\textbf{$ZZ$ interaction between the bus mode $n=15$ and target qubit $Q_t$}. The target qubit is set to several frequencies, shown in \textbf{a}. \textbf{b}, The predicted $ZZ_p/2\pi$ and experimental $ZZ_e/2\pi$ with corresponding color.}
    \label{fig:zz_0801}
\end{figure}

\begin{figure*}
    \centering
    \includegraphics[]{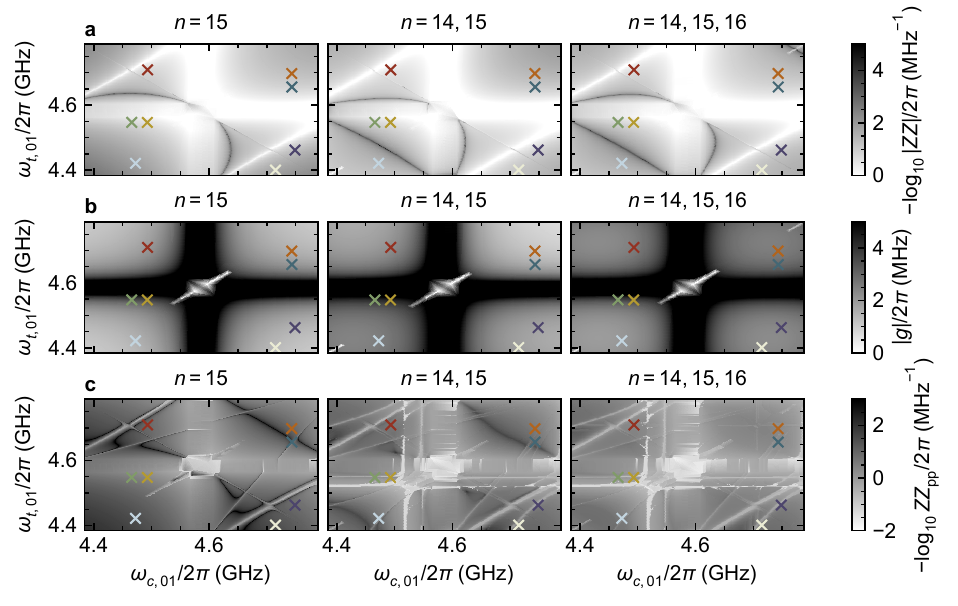}
    \includegraphics[]{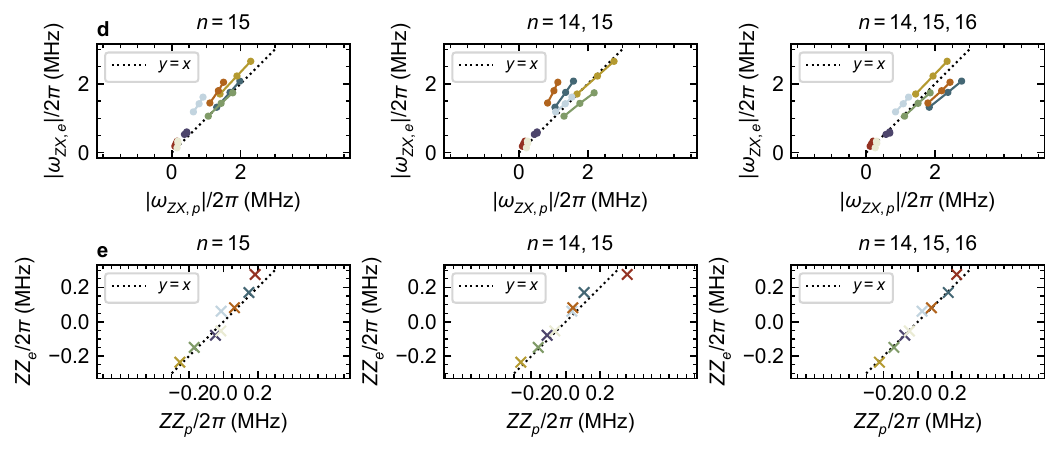}
    \caption{\textbf{Simulation results}. 
    The simulated \textbf{a} $ZZ$ interaction and \textbf{b} effective coupling strength $g$ between two qubits are considered for different coupling interaction, such as the $n=15$ (only the single mode considered), $n=14,15$ (two modes considered together), and $n=14,15,16$ (three modes considered together). The spectator effect from bus modes are shown in \textbf{c}, which is quantified by the peak-peak value of $ZZ$ strength when each bus mode is excited with one excitation or not. There are 2 (4, 8) $ZZ$ values in the $n=15$ ($n=14, 15$, $n=14, 15, 16$ case, and the difference between the maximum and minimum values are calculated. Several working frequencies of the control and target qubit are tested in experiments, labeled in different colors. The central white areas are ignored in numerical simulations since the frequencies of qubits and bus mode $n=15$ are too close to each other. Data outside the color bar range is represented in black or white. At these working frequencies, the \textbf{d} CR interaction strength and the \textbf{e} residual $ZZ$ interaction are measured and compared with simulation results. The prediction accuracy based three bus modes is better than the other two.}
    \label{fig:sim_g_zz}
\end{figure*}

The full Hamiltonian of our system takes the form of
\begin{equation}
\begin{aligned}
H&=\sum_{i=c, t} (\omega_ia_i^\dagger a_i+\frac{\alpha_i}2a_i^\dagger a_i^\dagger a_ia_i)+\sum_n\omega_na_n^\dagger a_n\\
&+\sum_ng_{c,n}\left(a_c^\dagger+a_c\right)\left(a_n^\dagger+a_n\right)\\
&+\sum_n(-1)^ng_{t,n}\left(a_t^\dagger+a_t\right)\left(a_n^\dagger+a_n\right),
\end{aligned}
\end{equation}
where the transmon is approximated by a Duffing oscillator. $\omega_c$ ($\omega_t$) is the transition frequency between the ground state $|0\rangle$ and the first-excited state $|1\rangle$, and the $\alpha_c$ ($\alpha_t$) is the anharmonicity. $a^\dagger$ and $a$ are the generation and annihilation operators of the corresponding bosonic mode. Here, the bus modes are media to coupling the two qubits. The two qubits are on different chips so that the direct coupling can be neglected.

For the most simplest case, coupling between a bus mode and a qubit is considered first. In experiments, we tune the bias of the control qubit making the frequency of $Q_c$ satisfying $\omega_c/2\pi<2$ GHz. Then the $ZZ$ interaction between the target qubit $Q_t$ and bus mode $n=15$ can be observed via exciting the bus mode $n=15$ and then measuring the difference of $\omega_t/2\pi$. Several $ZZ/2\pi$ values of different $\omega_t$ cases are tested and compared with their simulation predicted values based on the simple coupling system Hamiltonian $H=\omega_ta_t^\dagger a_t+\alpha_ta_t^\dagger a_t^\dagger a_ta_t/2+\omega_{15}a_{15}^\dagger a_{15}+g_{t, 15}\left(a_t^\dagger+a_t\right)\left(a_{15}^\dagger+a_{15}\right)$. These results are shown in FIG. \ref{fig:zz_0801}. The predicted $ZZ_p$ value is close to the experimental $ZZ_e$ value, indicating that the toy coupling model considered here are valid. However, the $ZZ$ interaction strength reaches the $1$ MHz level due to the large coupling strength $g_{t, 15}$, leading to a situation that such $ZZ$ interaction cannot be ignored in experiments. For instance, it is expected to make the bus mode $n=15$ on its ground state so that the derivative-removal-by-adiabatic gate (DRAG) method \cite{DRAG_exp, DRAG_theo} is employed to block the $\omega_{15}$ frequency component in the microwave pulse. Another instance is that the bus mode is regarded as a spectator of the two qubit gate. Meanwhile, the small differences between the experimental and predicted values are noticed, which indicates other bus modes also have some influences.

Then, the complexity of the total system increases with the number of bus modes considered. Given the frequency distance between the interaction region, the bus modes $n=15$, $n=14$, and $n=16$ have a higher priority to consider among all the bus modes. The effective coupling strength and residual $ZZ$ interaction strength between the control and target qubit are dominant in the coherent manipulation while the related bus modes are expected to keep on stable states, such as their ground states. According to this assumption, we make numerical simulations to acquire the interaction information via the eigenvalues and eigenstates of the system Hamiltonian. We vary the $\omega_c$ and $\omega_t$ with fixed $\alpha_c/2\pi=-240$ MHz and $\alpha_c/2\pi=-234.9$ MHz and calculate the dressed frequencies of qubits, effective two-qubit coupling strength $g$, and residual $ZZ$ strengths, as shown in FIG. \ref{fig:sim_g_zz}a and b. The anharmonicity here is an approximate estimation however it can show the coupling characteristics of the system. Residual $ZZ$ suppression region due to different modes can be found as the number of involved bus modes increasing. Less residual $ZZ$ interaction is desired to ensure the isolation of individual single-qubit operations. As for the coupling strength $g$, a proper strength value is needed to implement two-qubit entanglement, as fast as possible without causing higher-order effects. Meanwhile, the three bus modes considered here are near to the qubit working frequency. If the bus mode is excited, both $g$ and $ZZ$ are changed, which causes errors in the pulse operation, i.e. the spectator effect. We calculated the maximal $ZZ$ differences in the possible bus mode state with up to one excitation for each mode, as shown in FIG. \ref{fig:sim_g_zz}c. Together with the coherence of qubits, the working point to calibrate gate is determined.

The cross-resonance (CR) effect is the employed interaction to generate entanglement. The full CR Hamiltonian of our system can be written as
\begin{equation}
\begin{aligned}
    H_{\rm CR}=H&+\Omega_{c}\cos{(\omega_{t}t+\phi_c)}\left(a_c^\dagger+a_c\right)\\&+\Omega_t\cos{(\omega_{t}t+\phi_t)}\left(a_t^\dagger+a_t\right).
\end{aligned}
\end{equation}
Actually, according to the qubit layout, a leakage term $\Omega_n\cos(\omega_tt+\phi_n)\left(a_n^\dagger+a_n\right)$ is thought to exist simultaneously and to cause an extra CR effect on the target qubit (here, "control qubits" are the cavity modes). That is another reason why we need to make bus modes in stable states.

The CR-effect can be truncated into the calculation sub-space of control and target qubit only, in terms of
\begin{equation}
\begin{aligned}
    H_{\rm eff}&=\omega_{IX} IX+\omega_{IY} IY+\omega_{IZ} IZ\\&+\omega_{ZX} ZX+\omega_{ZY} ZY+\omega_{ZZ}ZZ,
\end{aligned} 
\end{equation}
with a rotating wave approximation (RWA) by $U=\exp{\left[-i\left(\omega_t ta_c^\dagger a_c+\omega_t ta_c^\dagger a_c\right)\right]}$. In this representation, the classical microwave crosstalk, the cancel drive, and CR-induced single qubit rotation are absorbed into the $IX$ and $IY$ term together. To confirm the CR effect strength, we take three different drive strengths and extract the CR strength $|\omega_{ZX,e}|$ parameters at different frequencies of qubits. Compared with the simulation predicted $|\omega_{ZX,p}|$ with three coupling mode cases, it is confirmed that the coupling strength calculated via single mode $n=15$ is weaker than the experimental result and is obvious that modeling the system with more bus modes is more convincing, as shown in FIG. \ref{fig:sim_g_zz}d. The results of $ZZ$ show the same characteristics in FIG. \ref{fig:sim_g_zz}e. The goodness of the predicted and experimental data indicates that multiple bus modes contribute to the coupling.

\subsection{Gate Calibration Method}

\begin{figure}[htbp]
    \centering
    \includegraphics[]{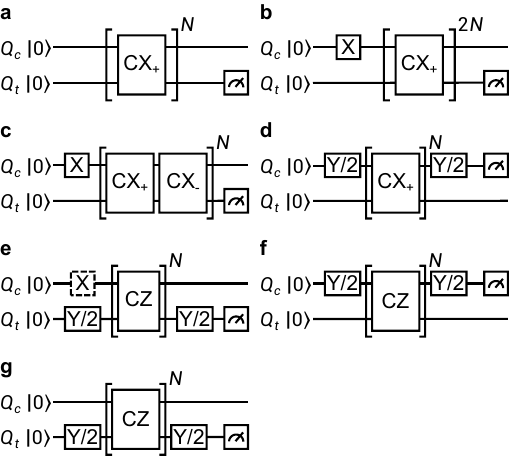}
    \caption{\textbf{Calibration circuits}. It is used for fine calibrations of CNOT and CZ gates.}
    \label{fig:cali_circ}
\end{figure}

Using the method in Ref \cite{CRparameters_sheldon}, we perform the CR Rabi experiments while the control qubit is on its $|0\rangle$ and $|1\rangle$ state respectively. On each control qubit state, we vary the driving time and use the quantum state tomography to get the final state Bloch vector $(X_{|0\rangle, |1\rangle}, Y_{|0\rangle, |1\rangle}, Z_{|0\rangle, |1\rangle})$. By fitting the result data with the Bloch equation model, we can extract the six parameters in Eq. (\ref{eq:CR_eff}) in the main text. 

For the CNOT case, our ultimate goal is to realize a two-qubit gate that the target qubit stays at the initial state (or rotates along the X-axis by an angle of $\pi$) when the control qubit is on its $|0\rangle$ (or $|1\rangle$) state. Firstly, we scan the CR pulse phase to find the phase at which the $ZX$ component is maximized while the $ZY$ component is zero. Then, for rough calibration, we scan the amplitude and phase of the active cancellation pulse on the target qubit to minimize the excitation on the target qubit when the control qubit is on $|0\rangle$.

Next, we apply the circuit sequences in FIG. \ref{fig:cali_circ} for fine calibration. To achieve a high-fidelity CNOT gate, we use the DRAG method to suppress the off-resonance driving-induced excitation on the control qubit and drive-induced ac-stark effect on the target qubit. During the fine calibration procedure, we apply an array of CNOT gates to amplify the error and thus find the best pulse parameters. Circuit a is used to calibrate the amplitude and phase of the cancellation pulse. Circuit b and c are used to calibrate the amplitude and the DRAG parameter of the CR drive. Circuit d is used for the control qubit frame changing calibration. So far, the gate is tuned to be $|0\rangle \langle 0|\otimes I+e^{i\varphi} |1\rangle \langle 1|\otimes X$. Finally, we cancel the drive-induced phase $\varphi$ with control qubit frame change and bring the gate to a standard CNOT.

The CZ gate is realized by driving the qubits simultaneously, at a frequency between the $|0\rangle \leftrightarrow |1\rangle$ and $|1\rangle \leftrightarrow |2\rangle$ transitions, causing state-dependent Stark shifts due to the state-dependent frequency detuning from the drive frequency. At certain driving amplitudes, we first scan the pulse duration and measure the phase accumulated on the target qubit when the control qubit is on its $|0\rangle$ or $|1\rangle$ state respectively. The gate duration is set to the time when the accumulated phase differs by $\pi$. Then, same as the CNOT gate calibration procedure, we also apply circuit calibration sequences for fine calibration. Circuit e is used for CZ gate duration calibration. Circuits f and g are used for qubits frame change calibration.

\subsection{XEB data}
We use the XEB method mentioned in \cite{blueprint_Martinis} to verify the fidelity of our single-bit gate and two-qubit gate fidelity.

For single-bit case, on each layer, we randomly choose a gate from $U(\theta)=\exp[-i\pi(X\cos\theta+Y\sin\theta)/4]$, where $\theta=0, \pi/4, 2\pi/4, \cdots, 7\pi/4$. We fit the decay curve of cross-entropy benchmarking fidelity with exponential function to get $p_{\rm ref}$, the average fidelity of the single-bit gate layer.

Then, we interleave the CNOT(CZ) gate into each layer and get the fidelity $p_{\rm int}$.
Finally, we calculate the CNOT(CZ) gate fidelity with $F_{\rm CNOT}=1-(D-1)(1-p_{\rm int}/{p_{\rm ref}})/D$, where $D$ is the system dimension(here, $D=4$).

For the CNOT(CZ) case, we take the data from 500 different measurements which lasts for 3.75 hours. Then, we use the bootstrap resampling method to resample the data for $10000$ times to calculate the deviation of the fidelity. We also choose different time windows to calculate the time-stability of the gate fidelity. (See main text)

\subsection{Initialization error}

\begin{figure}[htbp]
    \centering
    \includegraphics[]{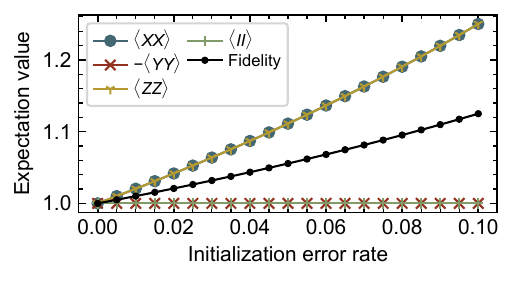}
    \caption{\textbf{Expectation value of stabilizers and Bell state fidelity as a function of initialization error} Simulation values of $\langle II \rangle$, $\langle XX \rangle$, $\langle -YY \rangle$, and$\langle ZZ \rangle$ at different initialization error when the conventional readout error correction method is applied.}
    \label{fig:thermal_GHZ}
\end{figure}

In the main text, we mentioned that our maximum CHSH correlation is slightly higher than the theoretical quantum limit. This might caused by the imperfection of state initialization. Usually, this can not be distinguished from measurement error in the single-qubit case. However, in the two-qubit case, with the help of a well-calibrated two-qubit entangling gate, these errors can be separated from each other. Here, we use a simple model to explain this in detail.

Let $P_{00}$, $P_{01}$, $P_{10}$, and $P_{11}$ be the populations on the two-qubit computational basis $|00\rangle$, $|01\rangle$, $|10\rangle$, and $|11\rangle$ respectively. Then, the outcome probability of the measurement $C_{00}$, $C_{01}$, $C_{10}$, and $C_{11}$ can be calculated as:
\begin{equation}
    \begin{pmatrix}
        C_{00}\\C_{01}\\C_{10}\\C_{11}
    \end{pmatrix}=M 
    \begin{pmatrix}
        P_{00}\\P_{01}\\P_{10}\\P_{11}
    \end{pmatrix},
\end{equation}
where $M$ is the $4 \times 4$ readout error matrix.

When the initialization error exists, the population on different basis can be written as:
\begin{equation}
    \begin{aligned}
        P_{00}&=(1-q_1)(1-q_2),\\
        P_{01}&=(1-q_1)q_2,\\
        P_{10}&=q_1(1-q_2),\\
        P_{11}&=q_1 q_2,\\
    \end{aligned}
\end{equation}
where $q_{1}$($q_{2}$) is the initialization error on qubit 1(2). We assume the initialization errors on two qubits are independent.

Conventionally \cite{shen2012correcting, nation2021scalable}, we apply single-qubit $\pi$-pulses to prepare the state into "$|00\rangle$", "$|01\rangle$", "$|10\rangle$", and "$|11\rangle$" and find out a $4 \times 4$ correction matrix $\Lambda$ which satisfies:

\begin{equation}
    \begin{split}
        \begin{pmatrix}
        1\\0\\0\\0
    \end{pmatrix} &=\Lambda M 
    \begin{pmatrix}
        P_{00}\\P_{01}\\P_{10}\\P_{11}
    \end{pmatrix}\\
        \begin{pmatrix}
        0\\1\\0\\0
    \end{pmatrix} &=\Lambda M 
    \begin{pmatrix}
        P_{01}\\P_{00}\\P_{11}\\P_{10}
    \end{pmatrix}\\
        \begin{pmatrix}
        0\\0\\1\\0
    \end{pmatrix} &=\Lambda M 
    \begin{pmatrix}
        P_{10}\\P_{11}\\P_{00}\\P_{01}
    \end{pmatrix}\\
        \begin{pmatrix}
        0\\0\\0\\1
    \end{pmatrix} &=\Lambda M 
    \begin{pmatrix}
        P_{11}\\P_{10}\\P_{01}\\P_{00}
    \end{pmatrix}.
    \end{split}
\end{equation}

However, this conventional method only works in the absence of initialization error (such as thermal excitation). Here, we use an example to demonstrate this. Consider the following situations: 1. no pulse is applied on the initial state, 2. a perfect CNOT gate is applied on the initial state. Ideally, the CNOT gate will not change the populations on each state if the initial state is $|00\rangle$, i.e.:
\begin{equation}
    \begin{split}
        \begin{pmatrix}
        1\\0\\0\\0
    \end{pmatrix} &=\Lambda M  
    \begin{pmatrix}
        P_{00}\\P_{01}\\P_{10}\\P_{11}
    \end{pmatrix}\\
        \begin{pmatrix}
        1\\0\\0\\0
    \end{pmatrix} &=\Lambda M
    \begin{pmatrix}
        P_{00}\\P_{01}\\P_{11}\\P_{10}
    \end{pmatrix}.
        \end{split}
\end{equation}
These equations are not consistent with each other. This shows the limitation of the conventional readout error correction method.

Meanwhile, this effect also provides us with a method to figure out the state preparation and measurement error separately. In the experiment, we apply single-qubit and two-qubit gates on the initial state to exchange the populations on each basis. By measuring the populations on each state without any correction, we get the following equations:
\begin{equation}
    \begin{split}
        \begin{pmatrix}
        0.9016\\0.0480\\0.0478\\0.0026
    \end{pmatrix} &=M 
    \begin{pmatrix}
        P_{00}\\P_{01}\\P_{10}\\P_{11}
    \end{pmatrix}\\
        \begin{pmatrix}
        0.0622\\0.8870\\0.0037\\0.0471
    \end{pmatrix} &= M 
    \begin{pmatrix}
        P_{01}\\P_{00}\\P_{11}\\P_{10}
    \end{pmatrix}\\
        \begin{pmatrix}
        0.0826\\0.0095\\0.8647\\0.0432
    \end{pmatrix} &= M 
    \begin{pmatrix}
        P_{10}\\P_{11}\\P_{00}\\P_{01}
    \end{pmatrix}\\
        \begin{pmatrix}
        0.0063\\0.0894\\0.0634\\0.8409
    \end{pmatrix} &= M 
    \begin{pmatrix}
        P_{11}\\P_{10}\\P_{01}\\P_{00}
    \end{pmatrix}\\
    \begin{pmatrix}
        0.8966\\0.0516\\0.0153\\0.0365
    \end{pmatrix} &= M 
    \begin{pmatrix}
        P_{00}\\P_{01}\\P_{11}\\P_{10}
    \end{pmatrix}\\
    \begin{pmatrix}
        0.1153\\0.0877\\0.0878\\0.7092
    \end{pmatrix} &= M 
    \begin{pmatrix}
        P_{01}\\P_{10}\\P_{11}\\P_{00}
    \end{pmatrix}.
    \end{split}
\end{equation}
There are 18 unknown parameters($M$ is a $4 \times 4$ real matrix, and $q_1$, $q_2$ are real numbers) and 24 equations. We use the least square method to optimize the parameters and get:
\begin{equation}
    \begin{split}
        M &=\begin{pmatrix}
            0.9676& 0.0238& 0.0466& 0.0454\\
            0.0115& 0.9536& 0.0121& 0.0566\\
            0.0172& 0.0053& 0.9271& 0.0634\\
            0.0037& 0.0172& 0.0141& 0.8345\\
        \end{pmatrix}
    \end{split}
\end{equation}
with $q_1=0.0399$ and $q_2=0.0373$. The initialization error on qubit1(2) is estimated to be $3.99 \%$($3.73 \%$). Since our CNOT gate fidelity exceeds $99 \%$, we believe it is capable of profiling these initialization errors. With these data, we can simulate the Bell-inequality test through the conventional readout error correction method and get the maximum CHSH correlation $S=2.94>2\sqrt{2}$. This correlation value is higher than the quantum limit and agrees with our experiment data well.

Since we do not utilize the higher energy level of our superconducting qubits, this method works for all the other quantum computing systems.

In FIG. \ref{fig:thermal_GHZ}, we show the simulation result of $\langle XX\rangle$, $\langle -YY\rangle$, and $\langle ZZ\rangle$ measurement result on Bell state ($|\Phi ^+\rangle =(|00\rangle +|11\rangle) / \sqrt{2} $) as a function of the initialization error on both qubits. This indicates that we will overestimate the Bell state fidelity if we use the conventional stabilizer method:
\begin{equation}
    F_{|\Phi ^+\rangle\langle \Phi ^+|} =\frac{\langle II\rangle+\langle XX\rangle+\langle -YY\rangle+\langle ZZ\rangle}{4}    
\end{equation}

\end{document}